\documentclass[journal]{IEEEtran}

\usepackage{amsmath, amsfonts}
\usepackage{bbm}
\usepackage{array}

\usepackage{algorithm}
\usepackage{algpseudocode}

\usepackage{enumitem}
\usepackage{graphicx}
\usepackage[caption=false,font=footnotesize]{subfig} % labelfont=sf,textfont=sf

\usepackage{url}
\usepackage{hyperref}
\usepackage{tikz}
\usepackage{multirow}
\usepackage{cite}
\def\BibTeX{{\rm B\kern-.05em{\sc i\kern-.025em b}\kern-.08em
T\kern-.1667em\lower.7ex\hbox{E}\kern-.125emX}}
\usepackage{amsmath}
\DeclareFontFamily{U}{mathx}{}
\DeclareFontShape{U}{mathx}{m}{n}{<-> mathx10}{}
\DeclareSymbolFont{mathx}{U}{mathx}{m}{n}
\DeclareMathAccent{\widehat}{0}{mathx}{"70}
\DeclareMathAccent{\widecheck}{0}{mathx}{"71}

\usepackage{amsthm}
\usepackage{dsfont}
\usepackage{seqsplit}
\usepackage{thmtools}
\usepackage{amssymb}
\usepackage[dvipsnames]{xcolor}

\def\bTheta{\boldsymbol{\Theta}}

\def\bUpsilon{\boldsymbol{\Upsilon}}

\def\bvarphi{\boldsymbol{\varphi}}
\def\bPhi{\boldsymbol{\Phi}}

\def\bOmega{\boldsymbol{\Omega}}

\def\mba{\mathbf{a}}

\def\mbe{\mathbf{e}}
\def\mbf{\mathbf{f}}

\def\mbh{\mathbf{h}}

\def\mbp{\mathbf{p}}

\def\mbu{\mathbf{u}}
\def\mbv{\mathbf{v}}
\def\mbw{\mathbf{w}}

\def\mbA{\mathbf{A}}

\def\mbD{\mathbf{D}}
\def\mbE{\mathbf{E}}

\def\mbI{\mathbf{I}}

\def\mbR{\mathbf{R}}

\def\mbU{\mathbf{U}}
\def\mbV{\mathbf{V}}
\def\mbW{\mathbf{W}}

\def\mbY{\mathbf{Y}}

\def\calA{\mathcal{A}}

\def\calC{\mathcal{C}}
\def\calD{\mathcal{D}}

\def\calH{\mathcal{H}}

\def\calL{\mathcal{L}}

\def\calN{\mathcal{N}}
\def\calO{\mathcal{O}}
\def\calP{\mathcal{P}}

\def\calS{\mathcal{S}}
\def\calT{\mathcal{T}}

\def\calX{\mathcal{X}}

\def\bzero{\boldsymbol{0}}

\newtheorem{remark}{Remark}

\def\T{\top}
\def\H{\mathrm{H}}

\newcommand{\figref}[1]{Fig.~\ref{#1}}
\newcommand{\secref}[1]{Section~\ref{#1}}
\newcommand{\tabref}[1]{Table~\ref{#1}}

\newcommand{\RX}{\text{RX}}
\newcommand{\TX}{\text{TX}}
\newcommand{\subc}{\text{s}}
\newcommand{\ant}{\text{a}}
\newcommand{\sym}{\text{sym}}
\newcommand{\back}{\mathsf{B}}
\newcommand{\FFT}{\text{2DFFT}}
\newcommand{\MUSIC}{\text{2DMUSIC}}
\newcommand{\JRAC}{\text{JRAC}}
\newcommand{\SRAE}{\text{SRAE}}

\usepackage{xifthen}
\newcommand{\Arng}[1][]{\ifthenelse{\isempty{#1}}{\bPhi_{\text{r}}}{\bPhi_{\text{r}, #1}}}
\newcommand{\brng}[1][]{\ifthenelse{\isempty{#1}}{\varphi_{\text{r}}}{\varphi_{\text{r}, #1}}}
\newcommand{\Aang}[1][]{\ifthenelse{\isempty{#1}}{\bPhi_{\angle}}{\bPhi_{\angle, #1}}}
\newcommand{\bang}[1][]{\ifthenelse{\isempty{#1}}{\bvarphi_{\angle}}{\bvarphi_{\angle, #1}}}
\newcommand{\Aall}{\bPhi}
\newcommand{\ball}{\bvarphi}

\title{Real-time Range-Angle Estimation and Tag Localization for Multi-static Backscatter Systems}
\author{Tara Esmaeilbeig$^{*}$, Kartik Patel$^{*}$, Traian E. Abrudan, John Kimionis, Eleftherios Kampianakis, Michael Eggleston
\thanks{$^{*}$ Two authors have equal contributions. All authors are with Nokia Bell Labs.
Email addresses: tara.esmaeilbeig@nokia.com,\{kartik.patel, traian.abrudan, ioannis.kimionis, lefteris.kampianakis, michael.eggleston\} @nokia-bell-labs.com}
}
\begin{document}
\maketitle 
\begin{abstract}
Multi-static backscatter networks (BNs) are strong candidates for joint communication and localization in the ambient IoT paradigm for 6G. Enabling real-time localization in large-scale multi-static deployments with thousands of devices require highly efficient algorithms for estimating key parameters such as range and angle of arrival (AoA), and for fusing these parameters into location estimates.
We propose two low-complexity algorithms, Joint Range-Angle Clustering (JRAC) and Stage-wise Range-Angle Estimation (SRAE). Both deliver range and angle estimation accuracy comparable to FFT- and subspace-based baselines while significantly reducing the computation. 
We then introduce two real-time localization algorithms that fuse the estimated ranges and AoAs: a maximum-likelihood (ML) method solved via gradient search and an iterative re-weighted least squares (IRLS) method. Both achieve localization accuracy comparable to ML-based brute force search albeit with far lower complexity. 
Experiments on a real-world large-scale multi-static testbed with 4 illuminators, 1 multi-antenna receiver, and 100 tags show that JRAC and SRAE reduce runtime by up to 40$\times$ and IRLS achieves up to 500$\times$ reduction over ML-based brute force search without degrading localization accuracy. 
The proposed methods achieve 3 m median localization error across all 100 tags in a sub-6GHz band with 40 MHz bandwidth. These results demonstrate that multi-static range-angle estimation and localization algorithms can make real-time, scalable backscatter localization practical for next-generation ambient IoT networks. 
% This makes them well-suited for real-time tag localization in large-scale BNs. Furthermore, 
%Experimental results demonstrate that both algorithms substantially reduce runtime while preserving localization performance, making them well-suited for real-time tag localization in  BNs.
\end{abstract}
\begin{IEEEkeywords}
Backscatter, integrated communications and sensing, %OFDM,
Joint range and angle estimation, %machine-type communications,
IoT.
\end{IEEEkeywords}
% \todo{1. Descriptive figure captions. \ate{Key message of what should be seen must be in the caption, but avoid huge captions}\\
% 3.\red{Observations: Semi-passive architecture impacting the performance. Call for a third state, the matched load, even for semi-passive tags.}}
\section{Introduction}
The ambient IoT paradigm envisioned for 6G demands large-scale connectivity and sensing solutions powered by ultra-low energy devices. Backscatter communication is a strong candidate to realize this vision~\cite{Zheng:AmbientIoT6G:2024,van_huynh_ambient_2018,buttAmbientIoTMissing2024}. While early commercial backscatter systems, based on  monostatic architectures~\cite{global_epc_epc_2008}, were constrained by short range and limited scalability, the recent advances in ambient and multi-static backscatter architectures overcome these limitations and support wide-area deployments while retaining the low-power benefits of backscatter devices~\cite{griffinCompleteLinkBudgets2009, wang_link_2008, kimionis2014increased, vougioukas_could_2016, alevizosMultistaticScatterRadio2018}. %Such multi-static BNs can be the key candidate to support the ambient IoT paradigm in 6G.

Prior art has proven the feasibility of multi-static and ambient backscatter networks (BNs) for communication with focus on data transmission and network scalability~\cite{abrudan2025next, patel2025scalability, 2014_KamKimTou, 2008_VanBleLei, bharadia_backfi_2015, kellogg_passive_2016, zhang_hitchhike_2016, zhang_freerider_2017, kellogg_wi-fi_2014, talla_lora_2017, wangMultiRiderEnablingMultiTag2024}
, with few examples exploring localization capabilities in sub-6GHz systems~\cite{skyvalakis2022elliptical, vestakisMultistaticNarrowbandLocalization2018, jiangWillowPracticalWiFi2024}.
Multi-static localization is critical for applications like asset tracking and context-aware IoT, manifesting a demand for scalable architectures and algorithms that jointly support communication and localization.
\begin{figure}[t]
\centering
	\input{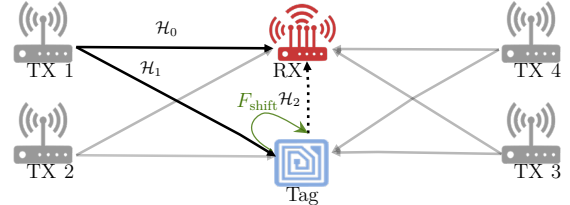}
    \vspace{-.1in}
     \caption{An illustration of a multi-static BN comprising 4 TXs, 1 multi-antenna RX, and a frequency-shifting tag: The tag shifts the carrier signal to an adjacent band to allow the RX process both signals simultaneously, and enable the bistatic range-angle estimation without external synchronization. Combining range-angle estimates from multiple TX-RX-pairs enable tag localization.}
    \label{fig:fig_1}
\end{figure}

In this work, we introduce a multi-static BN designed to natively support localization of tags. 
As illustrated in \figref{fig:fig_1}, the network consists of multiple single-antenna transmitters (TXs), multi-antenna receivers (RXs), and frequency-shifting backscatter tags~\cite{abrudan2025next}. By shifting and reflecting the incident carrier to an adjacent band, each tag enables the RX to process the direct and backscattered signals simultaneously. 
This dual-band processing \textit{(a)} isolates the backscatter signal from strong direct-path interference, and \textit{(b)} allows time-difference of arrival (TDoA)-based bistatic ranging using the direct TX signal as a timing without external synchronization between TXs and RXs. Furthermore, multi-antenna RXs support angle-of-arrival (AoA) estimation from  the backscatter signal.

We first develop bistatic range-angle estimation algorithms tailored for this architecture. 
Although range-angle estimation using wideband multi-antenna channels has been extensively studied~\cite{kim2018joint, jaafar2005joint}, scalable solutions suitable for networks with hundreds of tags remain an open challenge. To address this, we propose two methods: \textit{(a)}  Joint Range-Angle Clustering (JRAC) which is  a joint estimator that clusters dominant multipath components in a truncated range-angle heatmap. \textit{(b)} Stage-wise Range-Angle Estimation (SRAE) which is  a decoupled approach that estimates range and angle sequentially.
Compared to conventional approaches like 2D FFT~\cite{kim2018joint} and 2D MUSIC~\cite{jaafar2005joint} that incur prohibitive computational complexity,
%\ate{I expect 2-D FFT is less complex than JRAC, unless GPU is used to parallelize the heat map calculation}
% \te{please see table~\ref{tab:realtime_complexity_joint_range_aoa}, without parallelization the runtime of JRAC is less than  that of 2D-FFT.  The efficiency of JRAC comes from  truncation in the delay domain. The heatmap is calculated only on a window of  delays.}
JRAC and SRAE achieve near real-time operations without sacrificing accuracy.

Beyond estimation, we develop two real-time multi-static localization algorithms that fuse range and angle estimates: \textit{(a)} an ML-based localization solved using gradient ascent-based algorithm with line search, and \textit{(b)} an iterative re-weighted
least squares (IRLS) method that delivers comparable accuracy with dramatically lower computational costs. 

Finally, we validate our methods on a large-scale real-world testbed comprising 4 TXs, 1 RX, and 100 semi-passive tags deployed in a 20$\times$7 m indoor area. Despite heavy multipath conditions, our proposed algorithms achieve positioning accuracy comparable to baselines while reducing runtime by orders of magnitude. These results demonstrate the practicality of multi-static BNs for integrated communication and localization in next-generation ambient IoT networks. % As wireless systems transitions to 6G, energy efficiency emerges as a critical requirement for massive-scale and wide-area network deployments. Although conventional active radio technologies offer high performance, they are often prohibitively energy-intensive and costly for ultra-dense deployments involving sensors, wireless identification tags, or tracking devices. Consequently, such systems require alternative forms of machine-type communication to enable seamless integration into 6G-and-beyond networks. 

\subsection{Prior Art}\label{sec::related}
\subsubsection{Real-world Multi-static Localization Systems}
Recent BNs commonly adopt multi-static or ambient backscatter architectures for communication~\cite{2014_KamKimTou, 2008_VanBleLei, bharadia_backfi_2015, kellogg_passive_2016, zhang_hitchhike_2016, zhang_freerider_2017, kellogg_wi-fi_2014, talla_lora_2017, wangMultiRiderEnablingMultiTag2024}.
However, only a few works demonstrate real-world localization using multi-static architectures in sub-6 GHz bands~\cite{skyvalakis2022elliptical, vestakisMultistaticNarrowbandLocalization2018, jiangWillowPracticalWiFi2024}.
Work in ~\cite{skyvalakis2022elliptical} estimates direction-of-arrival and localizes RFID tags with a multistatic setup that requires a wired shared clock between radios. The latter, though effective for synchronization of co-located nodes, adds an element of deployment complexity for sparse, wide-area installations. Additionally, interference from multiple transmitters may become significant when TX and RX antennas are not co-linear, due to backscattering occuring in same channel. In contrast, our system leverages frequency-shifted backscatter, eliminating the need for external synchronization and reducing TX-induced interference.
Alternatively,~\cite{vestakisMultistaticNarrowbandLocalization2018} estimates bistatic ranges using received signal strength (RSS) measurements combined with particle-filter-based localization, which is appealing as a narrowband scheme. Our approach uses wideband signaling for illumination and phase-based ranging, which is more favorable in multipath environments~\cite{schroderInPhasePhasebased2022}, due to the full channel sounding characteristics. 
Finally,~\cite{jiangWillowPracticalWiFi2024} introduces digital TX-interference cancellation for in-band tag reflections. In our approach, by using frequency-shifting tags, we eliminate the need for such interference cancellation schemes, maximizing the receiver dynamic range independently for the illumination and backscatter signals, respectively. 

\subsubsection{Joint Range-Angle Estimation}
There are two main approaches for joint range-angle estimation. The first constructs a 2D range–angle heatmap and exploits structural assumptions to recover the Line of Sight (LoS) path. For example, compressive-sensing methods assume sparsity in the range-angle domain to search for the LoS component efficiently~\cite{xu2024joint,wu2022super}. However, operating in the sub‑6 GHz bands with less than 40 MHz bandwidth limits the sparsity in our setup, making such approaches less effective. 
The second class of methods, such as 2D FFT~\cite{kim2018joint} and 2D MUSIC~\cite{jaafar2005joint, kotaru2015spotfi}, compute the range–angle heatmap on a quantized grid and identify LoS peaks through max‑peak detection or iterative refinement, as in~\cite{jiangWillowPracticalWiFi2024}. Although accurate, these methods incur high computational cost due to dense grid evaluations. Our proposed JRAC and SRAE algorithms optimize this class of methods by truncating the grid search and using stage-wise range–angle estimation, resulting in substantial runtime reduction with minimal loss in accuracy.

\subsubsection{Localization Techniques}
We model errors in AoA measurements using directional statistics~\cite{directionalStat}. Since the angles are inherently periodic, they lie on a circular or spherical manifold rather than in Euclidean space. As a result, widely used statistical distributions are not representative of the angular errors. 
To the best of our knowledge, directional statistics were first applied to AoA-based localization using von Mises–Fisher distribution in~\cite{abrudan2016vmf}, and using von Mises distribution in~\cite{wang2012performance, naseri2018bayesian}. Subsequently, maximum likelihood (ML) based localization with Time-of-Arrival (ToA) measurements was introduced in~\cite{kuntal2022ml,perez2016blade}
for active-radio localization systems, which was later extended to incorporate joint range and angle measurements in~\cite{geng2021experimental}. In this paper, we extend this ML-based formulation to the multistatic setting and develop a gradient ascent algorithm with line search for efficiently estimating the tag position. 

Our second method, IRLS, is inspired by the pseudo-linear formulation for bi-static localization introduced in~\cite{chan1994simple}. Follow-up work applied similar ideas to active-radio settings with range-only measurements~\cite{ma2019tdoa, amiri2017asymptotically,varshney2025outlier} and AoA-only measurements~\cite{abrudan2024positioning}. 
In this paper, we extend IRLS to use both range and AoA measurements in a multi-static BN. Drawing from our experimental observations, we further refine the IRLS procedure to ensure robustness in practical deployments.    

Lastly, \cite{kanhereTargetLocalizationUsing2021, decarliRFIDRadarLocalization2013a} derive Cramer-Rao Lower Bound (CRLB) for multi-static localization but do not provide algorithms to achieve those limits. Moreover, temporal filtering techniques such as Kalman or particle filters can further improve localization performance~\cite{maromBistaticRadarTracking2023, vestakisMultistaticNarrowbandLocalization2018}. 
In this paper, we focus on raw localization performance, without exploring filtering, and assume that the localization accuracy can be further improved using subsequent filtering. 

% \ate{Stoica's work \cite{kuntal2022ml} is missing -- please add it and provide a compararison to it because it has ML and CRLB.}\te{it is a range only  work, I added the citation to~\cite{kuntal2022ml} above when explaining ML for ToA.}

\subsection{Summary of Contributions}
In summary, the contributions of this paper are as follows:
\begin{itemize}
\item We introduce a multi-static BN that natively supports tag localization along with communication, without any synchronization between TXs and RXs. The system uses the direct carrier signal as a timing reference for bistatic ranging and uses multi-antenna RXs for AoA estimation.
\item We present two efficient algorithms, JRAC and SRAE, for joint range--angle estimation in a bistatic backscatter system. We show these methods are computationally efficient without sacrificing the estimation accuracy.
\item We develop two localization algorithms that fuse bistatic range and angle. First, we formulate an ML estimator assuming Gaussian range errors and von~Mises--Fisher distributed angular errors, and solve it via gradient ascent with line search. Second, we introduce an IRLS-based method using pseudo-linear formulation. We show that both of these methods significantly reduce runtime while preserving overall localization performance.
\item We build a large-scale experimental testbed and collect two datasets: (a) with a 4-tag setup representing a simplified scenario, and (b) with a 100-tag setup representing a dense practical deployment. Using these datasets, we evaluate our estimation and localization methods, analyze their computation complexity (and runtime), and highlight key empirical observations on the challenging wireless environment data. 
\end{itemize}
\section{System Model}
%----------------------------------
\subsection{System Setup}
The multi-static BN consists of $N_\TX$ TXs and $N_\RX$ RXs located in $D$-dimensional space, $D\in\{2,3\}$. We denote the position of the $i$-th TX by $\mathbf{p}_{{\TX,i}}$, $j$-th RX by $\mathbf{p}_{{\RX,j}}$, and a backscatter tag by $\mathbf{p}$.

We denote by $d^{(i,j)}_0$ the distance between $i$-th TX and $j$-th RX. We further define the bistatic range as the distance traveled by a signal from $i$-th TX to $j$-th RX via the tag. Denoting the bistatic range by $d^{(i,j)}_{\back}$, we have \begin{equation}
    d^{(i,j)}_{\back} = ||\mbp-\mbp_{\TX,i}||+||\mbp-\mbp_{\RX,j}||.
\end{equation}
Let the angular position of the TX and the tag, i.e., AoA of the carrier and the backscatter signals, \textit{w.r.t.} the $j$-th receiver denoted by $(\theta_0^{(i,j)},\phi_0^{(i,j)})$, $(\theta_{2}^{(j)},\phi_{2}^{(j)})$, where $\theta_{0}^{(i,j)}, \theta_{2}^{(j)}\in[-\pi, \pi]$ denote the azimuthal angles and $\phi_0^{(i,j)},\phi_{2}^{(j)}\in[0,\pi]$ denote the elevation angles. Then, \begin{equation}\label{eq:GT-unitvector}
    \frac{\mbp-\mbp_{\RX, j}}{||\mbp-\mbp_{\RX, j}||} = \left[\sin \phi_{2}^{(j)} \cos \theta_{2}^{(j)},\sin \phi_{2}^{(j)} \sin \theta_{2}^{(j)},\cos \phi_2^{(j)}\right]^\T.
\end{equation}
In this work, we are interested in
\begin{enumerate}[label=(\alph*)]
\item estimating $\left(d^{(i,j)}_{\back},\theta_2^{(j)},\phi_2^{(j)}\right)$ from the carrier and the backscatter signals, and
\item calculating the tag position $\mbp$ using estimated $\left(d^{(i,j)}_{\back},\theta_2^{(j)},\phi_2^{(j)}\right)$ from all TXs and RXs.
\end{enumerate}
To simplify the notations, we avoid superscript $i,j$ when the indices of the TX and RX are clear from the context. 
\subsection{Signal Model}
In the following, we describe a bistatic signal model for an arbitrary $i$-th TX and $j$-th RX. Note that, collectively, the set of all TXs and RXs constitutes a \emph{multistatic} system, where each individual pair of TX, RX is a \emph{bistatic} system.

We consider the receivers are equipped uniformly-spaced $N_\ant$-element antenna array. %uniform linear array (ULA) in parallel to $+X$ direction with broadside direction along $+Y$ direction. Therefore, we focus on 
Denoting the position of $i$-th element of the antenna array by $[x_i, y_i, z_i]^\T$, we define the $(N_\ant \times 1)$ array response vector of the RX as \begin{align}
\mathbf{a}(\theta, \phi) &  = \Big[\exp\Big(\jmath \frac{2\pi}{\lambda} ((x_i-x_0) \cos\theta \sin\phi \nonumber\\
& + (y_i-y_0)  \sin\theta \sin\phi + (z_i-z_0) \cos\phi)\Big)\Big]_{i=0}^{N_\ant-1}. 
\end{align}

% We consider a uniform linear array (ULA) at the RXs with $N_\ant$ elements spaced at half-wavelength distances along $+X$ direction with the broadside direction along $+Y$ direction. Therefore, we focus on estimating only the range and azimuthal AoA. We, however, note that our signal model and the following range angle estimation and the positioning algorithms can be extended to planar arrays with azimuth and elevation AoA.

We consider the TXs transmit an OFDM-based wideband illumination signal, also called the \textit{carrier signal}, with the center frequency $F_{\text{c}}$ and $N_\subc$ equally-spaced subcarriers spanning the bandwidth $B$. 
The tag reflects the illumination signal by shifting the center frequency to $F_\text{c}+F_\text{shift},$ and modulating a packet on it. We call this reflected signal the \textit{backscatter signal}. In the following, we model the channels experienced by the carrier and the backscatter channels.

\subsubsection{Carrier Channel}
We consider $L_0$ propagation paths between the TX and RX, and denote the complex-valued coefficient, the bandwidth-normalized delay\footnote{If the delay is $t$, then the normalized delay is $\tau = Bt$ assuming the sampling rate of $B$. In practice, the RX sampling rate is higher than the bandwidth $B$. For the range and angle estimation, however, the sampling rate any more than the bandwidth does not provide further benefit. Hence, we assume the sampling rate of $B$.} and azimuth/elevation angles of $\ell$-th propagatikon path by $\alpha_{0,\ell}, \tau_{0,\ell}, \theta_{0,\ell}, \phi_{0,\ell}$, respectively.
We extend the single-antenna channel model from \cite{abrudan2025next} to write the  channel frequency response (CFR) of the TX-RX channel, called \textit{carrier channel}, for multi-antenna RX as 
\begin{align} 
{\boldsymbol{\calH}}_{0,n} = & \nonumber\\
 \sum_{\ell=0}^{L_0-1} \alpha_{0,\ell}&  \exp\left[-\jmath 2\pi \left(\frac{F_\text{c}}{B} + \frac{n}{N_\subc} \right) \tau_{0,\ell} \right] \mathbf{a}^{\T}\left( \theta_{0,\ell}, \phi_{0,\ell} \right),
\end{align}
where $n$ denotes the subcarrier index. 

After down-conversion to baseband and concatenating $\boldsymbol{\calH}_{0,n}$ for all subcarriers, we  have  $\boldsymbol{\calH}_{0}=[\boldsymbol{\calH}_{0,-N_\subc/2}^{\T},\ldots,\boldsymbol{\calH}_{0,N_\subc/2-1}^{\T}]^{\T} \in \mathbb{C}^{N_\subc \times N_\ant}$. Then, 
$\boldsymbol{\calH}_0$ satisfies the factorization
\begin{align}\label{eq::calH0}
&\boldsymbol{\calH}_0 = \mathbf{F}(\boldsymbol{\tau}_0)\mathbf{D}_0\mathbf{A}^\T(\boldsymbol{\theta}_0, \boldsymbol{\phi}_0),  % \\
% =&\underbrace{\big[ \mathbf{f}(\tau_{\back,0}), \ldots, \mathbf{f}(\tau_{\back,L-1}) \big]}_{N_\subc \times L}
% \underbrace{
% \begin{bmatrix}
% \alpha_{\back,0} & 0 & \cdots & 0 \\
% 0 & \ddots & \ddots & \vdots \\
% \vdots & \ddots & \ddots & 0 \\
% 0 & \cdots & 0 & \alpha_{\back, L-1}
% \end{bmatrix}
% }_{L \times L}
% \underbrace{\begin{bmatrix}
% \mathbf{a}^\T(\theta_0) \\
% \vdots \\
% \mathbf{a}^\T(\theta_{L-1})
% \end{bmatrix}}_{L \times N_\ant}, \nonumber 
\end{align}
where 
\begin{align}
\mathbf{F}(\boldsymbol{\tau})&= \left[\mathbf{f}(\tau_{0,0}, \ldots, \mathbf{f}(\tau_{0,L-1}) \right] \in \mathbb{C}^{N_\subc \times L}, \label{eq:F_tau}\\
\mathbf{f}(\tau)&=\left[e^{\jmath\frac{2\pi (N_\subc/2)}{N_\subc}\tau},\ldots,e^{\jmath\frac{-2\pi (N_\subc/2-1)}{N_\subc}\tau}\right] \in \mathbb{C}^{N_\subc \times 1}, \label{eq:f_tau}\\
\mbD_0&=\textrm{Diag}(\alpha_{0,0},\ldots,\alpha_{0,L-1})\in \mathbb{C}^{L \times L}, \label{eq:ch_coeff_0}\\
\mbA(\boldsymbol{\theta}_0,\boldsymbol{\phi}_0)&=\left[\mba(\theta_{0,0},\phi_{0,0}),\ldots,\mba(\theta_{0,L-1},\phi_{0,L-1})\right] \in \mathbb{C}^{N_\ant \times L}.\label{eq:A0}
\end{align}

\subsubsection{Backscatter Channel}

Similarly, we consider $L_1$ propagation paths between the TX and the tag, and denote the complex-valued coefficient, the bandwidth-normalized delay and AoA of $\ell$-th path by $\alpha_{1,\ell}, \tau_{1,\ell}, \theta_{1,\ell}, \phi_{1,\ell}$, respectively.
Then, the TX-Tag CFR is
\begin{equation} 
{\calH}_{1,n} = \sum_{\ell=0}^{L_1-1} \alpha_{1,\ell} \exp\left[-\jmath 2\pi \left(\frac{F_\text{c}}{B} + \frac{n}{N_\subc} \right) \tau_{1,\ell} \right],
\end{equation}
where $n$ is the subcarrier index. 

Similarly, we consider $L_2$ propagation paths between the tag and the RX, and denote the complex-valued coefficient, the bandwidth-normalized delay and AoA of $\ell$-th path by $\alpha_{2,\ell}, \tau_{2,\ell}, \theta_{2,\ell}, \phi_{2,\ell}$, respectively. Furthermore, the tag shifts the illumination signal by $F_\text{shift}$. Therefore, the Tag-RX CFR can be defined as 
\begin{align}
&{\boldsymbol{\calH}}_{2,n}
= \\
&\sum_{\ell=0}^{L_2-1} \alpha_{2,\ell} \exp\left[-\jmath 2\pi \left(\frac{F_\text{c}+ F_\text{shift}}{B} + \frac{n}{N_\subc} \right) \tau_{2,\ell} \right]\mathbf{a}^{\T}\left(\theta_{2,\ell}, \phi_{2,\ell} \right). \nonumber
\label{eq:multipath_ch2_discrete_freq_resp_tau_l} 
\end{align}

After  down  conversion to baseband, the CFR of the TX-Tag-RX channel, also called the \textit{backscatter channel}, is defined as 
\begin{align}
&\boldsymbol{\calH}_{\back,n}={\calH}_{1,n}    \boldsymbol{\mathcal{H}}_{2,n} \\ 
&=\sum_{\ell=0}^{L_1-1}\sum_{m=0}^{L_2-1}\Bigg[ \alpha_{1,\ell} \alpha_{2,m}  \exp\left(-\jmath \frac{2\pi n}{N_\subc}  (\tau_{1,\ell}+\tau_{2,m})\right)\nonumber\\ 
&\hspace{.4in}\exp\left(-\jmath 2\pi \left(\frac{F_\text{shift}}{B}  \right) \tau_{2,m}\right) \mathbf{a}^{\T}\left(\theta_{2,m}, \phi_{2,m} \right)\Bigg].
\label{eq:H_n}
\end{align}
Let a bistatic path, indexed by $\ell' = \ell + L_1 m$, denote the total signal path over $\ell$-th path in the TX-Tag channel and $m$-th path in the Tag-RX channel. Then, the effective channel coefficient, the normalized delay, and the AoA of $\ell'$-th bistatic path can be given by $\beta_{\ell'} = \alpha_{1,\ell}\alpha_{2,m}e^{-\jmath 2\pi F_\text{shift} \tau_{2,m}/B}$, $\tau_{\back,\ell'} = \tau_{1,\ell} + \tau_{2,m}$, and $\theta_{\back,\ell'} = \theta_{2,m}, \phi_{\back,\ell'} = \phi_{2,m}.$ % We highlight here that the AoA only depends on the path from Tag to RX.
As a result, \eqref{eq:H_n} can be simplified to
\begin{align}
&\boldsymbol{\calH}_{\back,n}=\sum_{\ell=0}^{L_1L_2-1} \beta_{\ell} \exp\left(-\jmath \frac{2\pi n}{N_\subc} \tau_{\back,\ell}\right) \mathbf{a}^{\T}\left(\theta_{\back,\ell},\phi_{\back,\ell} \right).
\end{align}
Concatenating $\boldsymbol{\calH}_{\back,n}$ for all subcarriers, we  have  $\boldsymbol{\calH}_{\back}=[\boldsymbol{\calH}_{\back,-N_\subc/2}^{\T},\ldots,\boldsymbol{\calH}_{\back,N_\subc/2-1}^{\T}]^{\T} \in \mathbb{C}^{N_\subc \times N_\ant}$. We define $L=L_1L_2$ as the total number of bistatic paths in the backscatter channel. Then, 
$\boldsymbol{\calH}_{\back}$ satisfies the factorization
\begin{align}\label{eq::calH}
&\boldsymbol{\calH}_{\back} = \mathbf{F}(\boldsymbol{\tau}_\back)\mathbf{D}_\back\mathbf{A}^\T(\boldsymbol{\theta}_\back, \boldsymbol{\phi}_\back),  % \\
% =&\underbrace{\big[ \mathbf{f}(\tau_{\back,0}), \ldots, \mathbf{f}(\tau_{\back,L-1}) \big]}_{N_\subc \times L}
% \underbrace{
% \begin{bmatrix}
% \alpha_{\back,0} & 0 & \cdots & 0 \\
% 0 & \ddots & \ddots & \vdots \\
% \vdots & \ddots & \ddots & 0 \\
% 0 & \cdots & 0 & \alpha_{\back, L-1}
% \end{bmatrix}
% }_{L \times L}
% \underbrace{\begin{bmatrix}
% \mathbf{a}^\T(\theta_0) \\
% \vdots \\
% \mathbf{a}^\T(\theta_{L-1})
% \end{bmatrix}}_{L \times N_\ant}, \nonumber 
\end{align}
where, 
% \mathbf{F}(\boldsymbol{\tau}_\back)&= \left[\mathbf{f}(\tau_{\back,0}, \ldots, \mathbf{f}(\tau_{\back,L-1}) \right] \in \mathbb{C}^{N_\subc \times L} \nonumber, \\
% \mathbf{f}(\tau)&=\left[e^{\jmath\frac{2\pi (N_\subc/2)}{N_\subc}\tau},\ldots,e^{\jmath\frac{-2\pi (N_\subc/2-1)}{N_\subc}\tau}\right] \in \mathbb{C}^{N_\subc \times 1} \nonumber, \\
$ \mbD_\back=\textrm{Diag}(\beta_0,\ldots,\beta_{L-1})\in \mathbb{C}^{L \times L}.$
% \mbA(\boldsymbol{\theta}_\back,\boldsymbol{\phi}_\back)&=\left[\mba(\theta_{\back,0},\phi_{\back,0}),\ldots,\mba(\theta_{\back,L-1},\phi_{\back,L-1})\right] \in \mathbb{C}^{N_\ant \times L}.
% \end{align}

In practice, the RX does not have access to the channel information directly, but processes the received signal to get the channel estimates~\cite{abrudan2025next}. We denote an estimate of the channel ${\boldsymbol\calH}$ by $\widehat{\boldsymbol\calH}$. 
Furthermore, the RX processes $N_{\sym}$ OFDM symbols to get a channel estimate. Let $\widehat{\boldsymbol{\calH}}[k]$ denote the estimate of $\boldsymbol{\calH}$ using $k$-th OFDM symbol. Therefore, the RX has access to $\widehat{\boldsymbol{\calH}}_{0}[k], \widehat{\boldsymbol{\calH}}_{\back}[k], \forall k\in \{1,...,N_\sym\}$ to estimate $\tau_{\back,0}, \theta_{\back,0}$ and $\phi_{\back,0}$. We assume the channels remain static for the duration of $N_\sym$ symbols. 

In next section, we discuss the joint range-angle estimation algorithms using these channel estimates. 
Without loss of generality, we subsequently assume the first path in all channels are the LoS path of the channels. Therefore, $\tau_i = \tau_{i,0}, \theta_i = \theta_{i,0}, \phi_i = \phi_{i,0}, i\in\{0,1,2,\back\}$. Furthermore, $\tau_i = Bd_i/c$. 
\section{Joint  Bistatic Range and angle Estimation}\label{sec:joint}
% \todo{Replace all equations with $\tau$ to follow $\tau = Bt$.}
For tag localization, we are interested in estimating the parameters $(\tau_{\back,0}, \theta_{\back,0}, \phi_{\back,0})$ of the LoS path in the backscatter channel corresponding to an arbitrary TX-RX pair. 

Notably, the bistatic range (or delay $\tau_{\back,0}$) estimation uses time difference of arrival (TDoA) method with the arrival of carrier signal as the reference~\cite{abrudan2025next}. This key design feature enables the ranging without requiring a separate clock sharing infrastructure between TXs and RXs in the network. 
Accordingly, let the TDoA estimate between the signal over the backscatter channel $\boldsymbol{\calH}_{\back}$ and the signal over the carrier channel $\boldsymbol{\calH}_0$ be denoted by $\Delta \widehat{\tau}$. Then, given the known TX-RX distance $d_{0}$, the estimate of the bistatic range is   
\begin{equation}
\widehat{d}_{B}=\frac{c}{B}  (\Delta \widehat{\tau} +\tau_{0,0}). 
\end{equation}
Estimating the TDoA $\Delta \widehat{\tau}$ involves estimating the time of arrival (ToA) over the backscatter channel  $\boldsymbol{\calH}_{\back}$ and the carrier channel  $\boldsymbol{\calH}_{0}$. Each ToA can be estimated using the channel impulse response (CIR)-based method as investigated in Section III.C of \cite{abrudan2025next}. 

In this work, in addition to estimating the bistatic range, we also estimate the AoA of the LoS path in the backscatter channel. This requires two ToA estimates, one from the carrier and one from the backscatter channel, and one AoA estimate from the backscatter channel.
To achieve this, we consider two approaches: (i) joint estimation of range and angle, and (ii) stage-wise estimation. The joint approach is employed by the baseline algorithms \textit{2D-FFT} and \textit{2D-MUSIC} described in \secref{sec::2DFFT} and \ref{sec::2DMUSIC}, as well as the proposed algorithm JRAC in \secref{sec::JRAC}. The stage-wise approach is used in the proposed algorithm SRAE in \secref{sec::SRAE}.

For the purposes of this work, we consider a ULA at the RXs and thus, estimate only azimuthal angles. Therefore, in this section, we assume $\phi_{\back,0} = \pi/2$. However, the proposed methods can be extended to enable estimating $\phi_{\back,0}$ with minimal modifications. 

Finally, we define the delay and angle grid as the ordered sets 
\begin{align}\label{eq::grid-sets}
\mathcal{T}&= \left\{\left. \frac{N_\subc i}{G_\tau}\right\vert i = 0,\ldots,G_\tau-1\right\} \quad\text{and} \\
% \mathcal{\bTheta}&=\left\{\left. j \frac{180}{G_{\theta}} - 90 \right| j=0,\ldots,G_{\theta}-1\right\}, \nonumber \\
\mathcal{\bTheta}&=\left\{\left. \sin^{-1}\left(\frac{2j}{G_{\theta}}\right)\right| j=-G_{\theta}/2,\ldots,G_{\theta}/2-1\right\},
\end{align} respectively, such that $|\mathcal{T}|=G_\tau$ and $|\mathcal{\bTheta}|=G_{\theta}$.

\subsection{2D FFT} \label{sec::2DFFT}
Let the 2D spectrum of $\widehat{\boldsymbol{\calH}}_{i}$ be \par \noindent \small
\begin{align}
&S^\FFT_{i}(\tau,\theta)= %\nonumber\\&
\sum_{k=1}^{N_\sym}\left|\sum_{m=-N_\subc/2}^{N_\subc/2-1}
\sum_{n=0}^{N_\ant-1}
[\boldsymbol{\widehat\calH}_{i}[k]]_{m+N_\subc/2+1,n+1}\,
e^{
\jmath 2\pi
\left(
\frac{\tau m}{N_\subc}
+
\frac{vn}{2}
\right)
}\right|,\label{eq:FFT-spectrum}
\end{align} \normalsize
where, $\tau \in \calT$,  $v=\sin \theta$ for $\theta\in\bTheta,$ and $ i\in\{0,\back\}$. % , and  
% \begin{align}
%     v'  = \begin{cases}
%     \sin \theta, & 0\leq \theta < \pi/2,\\
%     \sin \theta + 2, & -\pi/2 \leq \theta < 0
% \end{cases}.%v'/G_\theta),
% \end{align}
We define the set of $(\tau,\theta)$ where $|S^\FFT_{i}|$ has a peak above a threshold $T^\FFT_{\min}$ as \par \noindent \small
\begin{align}\label{eq::FFT-peakSearch}
& \calP^\FFT_i =\bigg \{(\tau_j,\theta_\ell) \in \calT\times\bTheta  \bigg| \; |S^\FFT_i(\tau_j,\theta_\ell)| > T^\FFT_{\min}, \nonumber \\
& |S^\FFT_i(\tau_j,\theta_{\ell-1})| <|S^\FFT_i(\tau_j,\theta_\ell)| > |S^\FFT_i(\tau_j,\theta_{\ell+1})|, \nonumber\\
& |S^\FFT_i(\tau_{j-1},\theta_\ell)|< |S^\FFT_i(\tau_j,\theta_\ell)| > |S^\FFT_i(\tau_{j+1},\theta_\ell)| \bigg\}.
\end{align} \normalsize
We then estimate the ToA of the carrier signal $\widehat{\tau}_{0,0}$ from $\calP^\FFT_0$ as 
\begin{align}
    \widehat{\tau}_{0,0} & = \underset{\tau: (\tau,\theta)\in \calP^\FFT_0}{\text{arg min}} \; \tau. % \nonumber\\
    % v_{0,0} & = \underset{(u_{0,0}, v)\in \calP^\FFT_0}{\text{arg min}} \; |S(u_{0,0},v)|.
\end{align}
Subsequently, we estimate the ToA and the AoA of the backscatter signal $(\widehat{\tau}_{\back,0}, \widehat{\theta}_{\back,0})$ from $\calP^\FFT_\back$ as 
\begin{align}
    \widehat{\tau}_{\back,0} & = \underset{\substack{(\tau,\theta)\in \calP^\FFT_\back \\ \tau > \widehat{\tau}_{0,0}}}{\text{min}} \; \tau, \nonumber\\
    \widehat{\theta}_{\back,0} & = \underset{\theta: (\widehat{\tau}_{\back,0}, \theta)\in \calP^\FFT_\back}{\text{arg max}} \; |S^\FFT_\back(\widehat{\tau}_{\back,0},\theta)|.
\end{align}

% corresponding to the lowest delay, i.e.
% attains the information of the range and angle of the first  path. If  $(u_0,v_0)$ are the frequency pairs satisfying~\eqref{eq::jointestfft} then, the estimated  delay and AoA of the first path is 
We use the resultant triplet $(\widehat{\tau}_{0,0}, \widehat{\tau}_{\back,0}, \widehat{\theta}_{\back,0})$ to estimate % to the associated delay and  angle estimates as follows: 
% \begin{align}
% \widehat{\tau}_{0,0} &=\frac{\widehat{u}_{0,0} N_\subc}{G_{\tau}}, \quad \quad \widehat{\tau}_{\back,0} =\frac{\widehat{u}_{\back,0} N_\subc}{G_{\tau}}, \nonumber \\
% \widehat{\theta}_\back & = \widehat{\theta}_{\back,0} = \begin{cases}
%     \sin^{-1}\left(\frac{2\widehat{v}_0}{G_\theta}\right), & 0\leq \widehat{v}_0 < G_\theta/2 -1,\\
%     \sin^{-1}\left(\frac{2\widehat{v}_0}{G_\theta}-2\right), & G_\theta/2 \leq \widehat{v}_0 < G_\theta
% \end{cases}.%v'/G_\theta),
% \end{align}
% Therefore, 
the bistatic range and the AoA as follows: 
\begin{align}\label{eq::FFT-range-est}
    \widehat{d}_\back = \frac{c}{B} (\tau_{0,0} + \widehat{\tau}_{\back,0} - \widehat{\tau}_{0,0}), \quad\text{and} \quad \widehat{\theta}_\back = \widehat{\theta}_{\back,0}.
\end{align}
Finally, $(\widehat{d}_\back,\widehat{\theta}_\back)$ presents the resultant range-angle pair associated to the TX-Tag-RX path. 
The computational complexity of the 2D FFT method is $\mathcal{O}(G_\tau G_\theta \log_2 (G_\tau G_\theta))$.

\subsection{2D MUSIC}\label{sec::2DMUSIC}
Two-dimensional MUSIC algorithm is an extension of MUSIC algorithm to jointly estimate ToA and AoA~\cite{jaafar2005joint,kotaru2015spotfi}. We use this algorithm to estimate $(\widehat{d}_\back, \widehat{\theta}_B)$ for a bistatic system.

% We leverage  the fact that having many OFDM subcarriers enables us to get a CSI measurement on each subcarrier. Therefore, the number of sensors can be expanded to be equal to the product of the number of subcarriers and the number of receiver  antennas~\cite{jaafar2005joint,kotaru2015spotfi}. Specifically, consider the sensor array comprising of all the subcarriers at all the antennas. 

Let the measurement vector $\widehat{\mbh}_i[k]=\textrm{vec}(\widehat{\boldsymbol{\calH}}_i[k]) \in \mathbb{C}^{N_\subc N_\ant \times 1}$ be a stack of the channel estimates of all the subcarriers at all the antennas, for $i=\{0,\back\}$ and $k=\{1,\ldots,N_\sym\}$. 
Then, from \eqref{eq::calH0} and \eqref{eq::calH}, we have% the collected observations from $k$-th OFDM symbol periods as 
\begin{align}\label{eq::vec-h}
&\widehat{\mbh}_i[k]=\textrm{vec}(\boldsymbol{\calH}_{i}[k])+\mbv_i[k]\\
&\overset{(a)}{=} \textrm{vec}(\mathbf{F}(\boldsymbol{\tau}_i)\mathbf{D}_i\mathbf{A}^\T(\boldsymbol{\theta}_i)  )+\mbv_i[k] %\nonumber\\&
=\mbU(\boldsymbol{\tau}_i,\boldsymbol{\theta}_i) \boldsymbol{\gamma}_i[k]+\mbv_i[k],\nonumber
\end{align}
where $\mbv_i[k]$ denotes channel estimation error, $(a)$ assumes static channel over $N_\sym$ symbols, 
% for $n=1,\ldots,N_{\sym}$ and 
$\boldsymbol{\gamma}_0=\boldsymbol{\alpha}_0$, $\boldsymbol{\gamma}_\back = \boldsymbol{\beta}_\back$, and \begin{align}
\mbU(\boldsymbol{\tau},\boldsymbol{\theta})= \left[\mba(\theta_0)\otimes \mbf(\tau_0),\ldots,\mba(\theta_{L-1})\otimes \mbf(\tau_{L-1})\right].
\end{align}
The collected
observations from all $N_{\sym}$ OFDM symbols is given by
\begin{align}
\mbY_{i}&=\left[\widehat{\mbh}_{i}[1],\ldots,\widehat{\mbh}_{i}[N_{\sym}]\right]  =\mbU(\boldsymbol{\tau}_i,\boldsymbol{\theta}_i)(\boldsymbol{\gamma}_i \otimes \mathbf{1}^\T_{N_\sym})+\mbV_i  ,
\end{align}
where $\mbV_i=\left[\mbv_i[1],\ldots,\mbv_i[N_{\sym}]\right]\in \mathbb{C}^{N_\subc N_\ant \times N_{\sym} }$.

The 2D MUSIC algorithm is based on  decomposition of the sample  covariance matrix, $\mbR_{\mbY_i}=\frac{1}{N_{\sym}}\mbY_{i}\mbY_{i}^{\H}$, into the signal subspace $\mbE_\text{S}$ and noise subspace $\mbE_{\text{N}}$ as 
\begin{equation}
\mbR_{\mbY_i}=\mbE_{\text{S},i} \boldsymbol{\Lambda}_{\text{S},i} \mbE_{\text{S},i}^{\H}+ \mbE_{\text{N},i} \boldsymbol{\Lambda}_{\text{N},i}\mbE_{\text{N},i}^{\H},
\end{equation}
where $\boldsymbol{\Lambda}_{\text{S},i}$ and $\boldsymbol{\Lambda}_{\text{N},i}$ are diagonal matrices with the eigenvalues of the corresponding subspaces.  
Assuming the number of signal space $K_i$, $\boldsymbol{\Lambda}_{\text{S},i}$ and $ \mbE_{\text{S},i}$ are determined by using the $K_i$-largest eigenvalues and associated eigenvectors of $\mbR_{\mbY_i}$, whereas  $\boldsymbol{\Lambda}_{\text{N},i}$ and $\mbE_{\text{N},i}$ are determined using the remaining $N_\subc N_\ant-K_i$ eigenvalues and the associated eigenvectors. 
In practice, the value of $K_i$'s is typically unknown and various criterion such as Minimum Description Length (MDL), Akaike Information Criterion (AIC) are used for estimating $K_i$'s.

The orthogonality between $\mbu(\tau,\theta)=\mba(\theta) \otimes \mbf(\tau)$  and the noise subspace is measured as $\|\mbu(\tau,\theta)^{\H}\mbE_{N}\|_2^2$, and its inverse forms the pseudo-spectrum of 2D MUSIC as  
\begin{align} \label{eq:music-spec}
&S_i^{\MUSIC}(\tau,\theta)=%\frac{1}
({\mbu(\tau,\theta)^{\H}\mbE_{\text{N},i}\mbE_{\text{N},i}^{\H}\mbu(\tau,\theta)})^{-1},%\; \text{where,}\\
\end{align}
for $(\tau,\theta) \in \calT \times \bTheta$ and $i\in\{0,\back\}$. 

Finally, to estimate $(\widehat{d}_\back,\widehat{\theta}_\back)$ from $S^\MUSIC_i, i\in\{0,\back\}$, the procedure described in \eqref{eq::FFT-peakSearch}-\eqref{eq::FFT-range-est} is followed, albeit with the quantities $T^\MUSIC_{\min}$ and $\calP_i^\MUSIC$. 
%We summarize our implementation of 2D MUSIC algorithm in Algorithm~\ref{alg::alg-2DMUSIC}.

The computational complexity of the 2D-MUSIC algorithm is dominated by three main components. First, the eigenvalue decomposition (EVD) of the sample covariance matrix $\mbR_{\mbY_i}$ incurs a cost of $\mathcal{O}(N_\ant^3 N_\subc^3)$. Second, the matrix multiplication in~\eqref{eq:music-spec} involving matrices of size $N_\ant N_\subc \times 1$ and $N_\ant N_\subc\times (N_\ant N_\subc-K_i)$ contributes to the complexity of $\mathcal{O}\left(N_\ant^{2} N_\subc^{2}\right)$. Finally, the pseudo-spectrum calculation, which evaluates the spectrum over $G_\tau G_\theta$ grid points, scales the computational complexity linearly with the number of grid points. Therefore, the computational complexity of the 2D MUSIC algorithm is $\mathcal{O}(G_\tau G_\theta N_\ant^2 N_\subc^2+N_\ant^3 N_\subc^3)$. 
This computational cost may pose significant limitations for practical implementations. Therefore, in the following sections, we investigate alternative approaches for joint range and angle estimation that offer improved computational efficiency.
\subsection{Stage-wise Range-Angle Estimation (SRAE)}\label{sec::SRAE}
This algorithm operates in two stages:
\subsubsection{Stage 1 (Delay estimation)}
To estimate the LoS delay $\widehat{\tau}_i$, we apply the (one-dimensional) MUSIC algorithm in the delay domain. % the estimated channels $\boldsymbol{\calH}_i$ for $i \in \{0,\back\}$. 
The sample covariance matrix is defined as 
\begin{align} \label{eq::cov_i}
\mathbf{R}_{\widehat{\boldsymbol{\calH}}_i} = \frac{1}{N_{\sym}} \sum_{k=1}^{N_{\sym}} \widehat{\boldsymbol{\calH}}_i [k]\widehat{\boldsymbol{\calH}}_i^{\H}[k] \;\in\; \mathbb{C}^{N_\subc \times N_\subc}.
\end{align}
The covariance matrix is then decomposed into the signal subspace $\mbE_\text{S}$ and noise subspace $\mbE_{\text{N}}$ using the MDL criterion.
The MUSIC spectrum is then calculated as 
\begin{align}\label{eq::SSRAE}
S_i^\SRAE(\tau)=({\mbf(\tau)^{\H}\mbE_{\text{N},i}\mbE_{\text{N},i}^{\H}\mbf(\tau)})^{-1}.
\end{align}
Then, the peak search is performed on the delay-spectrums of the carrier and  backscatter channels, such that \par \noindent \small
 \begin{align}
\widehat{\tau}_{0,0}&={\min} \bigg \{\tau_j \in \calT \bigg |  S_0^\SRAE(\tau_j)>T^\SRAE_{\min},  \nonumber\\
& \hspace{.3in}S_0^\SRAE(\tau_{j-1}) < S_0^\SRAE(\tau_j) > S_0^\SRAE(\tau_{j+1})  \bigg\}, \\
\widehat{\tau}_{\back,0}&={\min} \bigg \{\tau_j \in \calT \bigg |  \tau_j > \widehat{\tau}_{0,0}, S_\back^\SRAE(\tau_j)>T^\SRAE_{\min},  \nonumber\\
& \hspace{.3in}S_\back^\SRAE(\tau_{j-1}) < S_\back^\SRAE(\tau_j) > S_\back^\SRAE(\tau_{j+1})  \bigg\}. \label{eq:music-spec2}
\end{align} \normalsize
Using the delay estimates $\widehat{\tau}_{0,0},\widehat{\tau}_{\back,0}$, the bistatic range estimate can be determined using \eqref{eq::FFT-range-est}.  
%\figref{fig:alg1-range} gives an example of $S_B^\SRAE(\tau)$ for the channel setup described in \figref{fig:alg-comparison-geometry}.

% %-----------------------------
% \begin{algorithm}[tb] 
% \caption{Stage-wise Range-Angle Estimation (SRAE)\label{alg::SRAE}}
% % \scalebox{0.87}{%
% % \begin{minipage}{1.15\linewidth}
% \begin{algorithmic}[1] 
% \Statex\textbf{Input} $\boldsymbol{\calH}_0[k], \boldsymbol{\calH}_\back[k], \forall k\in\{1,\ldots,N_\sym\}$, $\tau_{0,0}$, $T^\SRAE_{\textrm{min}}$, $\mathcal{T}$, $\mathcal{\bTheta}$.
% \State \textbf{Stage 1: } Estimate  $\widehat{\tau}_{i,0}$ uing \eqref{eq::cov_i}-\eqref{eq:music-spec2} $\forall i \in \{0,\back\}$. 
% \State \textbf{Stage 2: }Follow \eqref{eq:X_B}-\eqref{eq:music-spec1} to estimate $\widehat{\theta}_{\back}$.
% \State Calculate the bistatic range as $\widehat{d}_\back=c/B( \widehat{\tau}_{\back,0} -\widehat{\tau}_{0,0} +\tau_{0,0})$.
% \Statex\textbf{Output} The estimated bistatic range and angle as  $(\widehat{d}_{\back}, \widehat{\theta}_\back)$.
% \end{algorithmic}
% \end{algorithm}

\subsubsection{Stage 2 (AoA  estimation)}
The complex gains of the channel taps associated to the LoS path for all antennas can be retrieved using the channel estimate $\widehat{\boldsymbol{\calH}}_\back$. These taps associated to the delay $\widehat{\tau}_{\back,0}$ is given by
\begin{align}\label{eq:X_B}
\widehat{\boldsymbol{\calX}}_\back[k]=\mbf(\widehat{\tau}_{\back,0})^{\dagger}\widehat{\boldsymbol{\calH}}_\back[k],\  k\in\{1,\ldots,N_\sym\},
\end{align}

The  AoA estimation of the LoS path of the backscatter channel, $\widehat{\theta}_{\back}$,  is obtained based on the relative phases of $\widehat{\boldsymbol{\calX}}_\back$, again by applying either beamforming or subspace-based methods. 
In this instance, we run another round of MUSIC in a spatial domain. The relevant sample covariance 
$
\mbR_{\calX_\back}=\frac{1}{N_{\sym}}\sum_{k=1}^{N_{\sym}}   \boldsymbol{{\calX}}_\back^{\H}[k] \boldsymbol{{\calX}}_\back[k] ,  
$
is decomposed into the signal subspace $\mbE_{\text{S},\back}$ and noise subspace denoted by $\mbE_{\text{N},\back}$. Note that, given there is only single-delay (and hence, single path) in \eqref{eq:X_B}, the size of the signal subspace is 1. 
The orthogonality between $\mba(\theta)$  and the noise subspace is measured as $\|\mba(\theta)^{\H}\mbE_\text{N}\|_2^2$, the inverse of which forms the MUSIC spectrum in spatial domain. This spectrum is searched for the peaks as
\begin{equation} \label{eq:music-spec1}
\widehat{\theta}_{\back}=\underset{\theta\in \bTheta}{\textrm{argmax}}\;\frac{1}{\mba(\theta)^{\H}\mbE_{\text{N},\back}\mbE_{\text{N},\back}^{\H}\mba(\theta)}.
\end{equation}
%\figref{fig:alg1-aoa} gives an example of the MUSIC spectrum in the spatial domain, corresponding to the channel taps associated to the estimated delay in \figref{fig:alg1-range}.
% We summarize the SRAE algorithm in Algorithm~\ref{alg::SRAE}.

SRAE method is equivalent to applying two one-dimensional MUSIC algorithms. The computational complexity of range estimation is $\mathcal{O}(G_{\tau} N_{\text{s}}^{3})$, while the complexity of the AoA estimation is $\mathcal{O}(G_{\theta} N_{\text{a}}^{3})$. Therefore, the overall computational complexity is $\mathcal{O}(G_{\tau} N_{\text{s}}^{3} + G_{\theta} N_{\text{a}}^{3})$. 

With the SRAE algorithm, the accuracy of delay estimation affects the accuracy of the estimated taps, and consequently, of the AoA estimation.
Hence, this stage-wise approach remains sub-optimal, in fact, irrespective of the order of estimating the parameters. Thus, we next introduce a joint estimation method based on clustering on the joint range-angle heatmap.
%------------------------------------------------
\subsection{Joint Range-Angle Clustering (JRAC)}\label{sec::JRAC}
The JRAC method follows a four-step process: First, similar to SRAE, it identifies the appropriate search space in delay domain. Second, it estimates a 2D heatmap for the estimated delay space and the entire spatial domain. Third, it clusters various channel taps. Finally, it selects the cluster with the lowest delay to retrieve the delay and angle corresponding to the LoS path. In the following, we formally describe each of these steps in detail. The outline of our algorithm is given in Algorithm~\ref{alg::JRAC}. 

\subsubsection{Step 1 (Truncate the search space in delay-domain)}
% In general, with an OFDM signal with $N_\subc$ subcarriers and $B$ total bandwidth, the maximum delay that can be estimated without ambiguity is $N_\subc/B$. 
We first truncate the possible delay space using 1D MUSIC in the delay domain.
Specifically, we use $S_i^\SRAE, i\in\{0,\back\}$ from~\eqref{eq::SSRAE}, and the threshold $T_{\min}^\JRAC$ to determine the truncated delay space as 
\begin{align}
    \calT_i = \left\{\tau \in\calT | S_i^\SRAE(\tau) \geq T_{\min}^\JRAC\right\}, i \in \{0,\back \}.
\end{align}

\subsubsection{Step 2 (The range-angle heatmaps and clipping)}
We define the 2D heatmap associated to the estimated channels $\widehat{\boldsymbol{\calH}}_i$ as \par \noindent \small
\begin{align}\label{eq::jointest1}
S_i^\JRAC(\tau,\theta)=
\sum_{k=1}^{N_{\sym}}\left\|\mbf(\tau)^{\H}\widehat{\boldsymbol{\mathcal{H}}}_i[k]\mathbf{a}^*(\theta)\right\|^2,\  {\genfrac{}{}{0pt}{0}{\tau \in \calT_i, \theta \in \bTheta,}{i \in \{0,\back\}.}}
\end{align} \normalsize
%An example of such heatmap is depicted in \figref{fig:alg2-heatmap}.
Note that, this heatmap is similar to the oversampled 2D FFT spectrum described in \eqref{eq:FFT-spectrum}, albeit with a truncated delay space with reduced computational cost. 
Hence, it is possible to estimate the bistatic range and angle using equations \eqref{eq::FFT-peakSearch}-\eqref{eq::FFT-range-est}. However, in the following, we introduce a clustering step which can provide a robust estimate in an heavily multipath environment.

Let $T^\JRAC_{\max}$ denote a lower clipping threshold and $s_{\max,i} = \max_{\tau\in\calT_i,\theta\in\bTheta} S_{i}^{\JRAC}(\tau,\theta)$ denote the maximum value of the heatmaps for $i\in\{0,\back\}$. We define a clipped normalized spectrum as \par \noindent \small
\begin{equation}
    \bar{S}_i^\JRAC(\tau,\theta) = S_i^\JRAC(\tau, \theta)\mathbf{1}\{S_i^\JRAC(\tau, \theta)\geq T_{\max}^\JRAC s_{\max,i}\}.
\end{equation} \normalsize
Notably, $\bar{S}_i^{\JRAC}, i\in\{0,\back\}$ have islands of clusters.
\begin{algorithm}[tb]
\caption{Joint Range-Angle Clustering (JRAC)\label{alg::JRAC}}
\scalebox{0.8}{%
  \begin{minipage}{0.95\linewidth}
\begin{algorithmic}[1]
\Statex\textbf{Input} $\widehat{\boldsymbol{\calH}}_i[k], \tau_{0,0}, \calT, \mathcal{\bTheta}$, Thresholds $T^\JRAC_{\text{min}}$ and $T^\JRAC_{\text{max}}$.
\For{$i\in\{0,\back\}$}
\Statex \textbf{Step 1: Truncate the search space in delay-domain}
\State Calculate  $S_0^\SRAE(\tau)$  and $S_\back ^\SRAE(\tau)$ according to~\eqref{eq::SSRAE}. 
\State Truncate  the  delay search space $\calT$ to get $\calT_0, \calT_{\back}$ as 
$$\calT_0= \bigg\{ \tau \in \calT \,\big|\, 
 S_0^\SRAE(\tau)   \ge T_{\min} \bigg\},$$
$$\calT_{\back}=\bigg\{ \tau \in \calT \,\big|\, 
 S_\back^\SRAE(\tau)  \ge T_{\min} \bigg\}.$$
\Statex \textbf{Step 2: The range-angle heatmaps and clipping}
\State Get the 2D heatmap $S_i^\JRAC(\tau,\theta), \forall\tau \in \calT_i, \theta \in \bTheta$ using \eqref{eq::jointest1}. 
\State Find the maxima of the heatmap $s_{\max,i} = \underset{\tau,\;\theta}{\max}\; S_i^\JRAC(\tau,\theta)$, then clip the $S_i^\JRAC$ such that 
    $$\bar{S}_i^\JRAC(\tau,\theta) = S_i^\JRAC(\tau, \theta)\mathbf{1}\{S_i^\JRAC(\tau, \theta)\geq T_{\max}^\JRAC s_{\max,i}\}.$$
\Statex \textbf{Step 3.a: Clustering in delay-domain}
\State Calculate $ \bar{s}_i(\tau) = \sum_{\ell = 1}^{|\bTheta|} \bar{S}_i^\JRAC(\tau,\theta_\ell),\; \forall \tau \in\calT_i,$.
\State Determine boundaries of the delay-domain clusters by \label{alg:line:boundary} $$t_{i,j} = \begin{cases}
        0, j = 0 \ \text{ or } \  j = |\calT_i|+1,\\
        1, \bar{s}_i(\tau_j) \neq 0,\\
        0, \bar{s}_i(\tau_j) = 0,
    \end{cases}, j \in \{0,\ldots,|\calT_i|+1\},$$
\State $b_{i,j}  = t_{i,j+1} - t_{i,j}, j \in \{0,\ldots,|\calT_i|\}.$
\State Determine the range-clusters $\calD_{i,m}$ using \eqref{eq:delay-cluster}.\label{alg:line:cluster}
\Statex \textbf{Step 3.b: Clustering in angle-domain}
\State Calculate 
$\bar{s}_i(\theta_\ell)
=
\sum_{j=1}^{|\calT_i|}
\bar{s}_i(\tau_j,\theta_\ell), \forall \theta_j \in \bTheta$.
\State Repeat Steps \ref{alg:line:boundary}-\ref{alg:line:cluster} to detect the angle-clusters. Denote $n$-th cluster as $\calA_{i,n}$. 
\Statex \textbf{Step 4: Estimation}
\State Let $k$-th cluster $\mathcal{C}_{i,k}= \mathcal{D}_{i,m} \times \mathcal{A}_{i,n}$. 
\State Determine the local maxima within each cluster as $$ (\widehat{\tau}_{i,k},\widehat{\theta}_{i,k}) = \underset{(j,\ell)\in\calC_{i,k}}{\text{arg max}} \bar{S}_i^\JRAC(\tau_j,\theta_\ell). $$
\State Find the LoS cluster indexed by $k_i^*={\text{arg min}_k}  \widehat{\tau}_{i,k}$.
\EndFor
\State Calculate $ \widehat{\tau}_{0,0} = \tau_{0,k_0^*}, \widehat{\tau}_{\back,0} = \tau_{\back,k_\back^*}, \widehat{\theta}_{\back, 0} = \theta_{\back,k_\back^*}.$
\State Repeat steps 4-9 on $S_0^\JRAC(\theta,\tau)$ to estimate the TX-RX delay $\widehat{\tau}_{0,0}$.
\State Calculate the LoS bistatic range and angle as  $$    \widehat{d}_\back = \frac{c}{B} (\tau_{0,0} + \widehat{\tau}_{\back,0} - \widehat{\tau}_{0,0}) \quad\text{and} \quad \widehat{\theta}_\back = \widehat{\theta}_{\back,0}.
$$
\Statex\textbf{Output} The  estimated bistatic range and angle as $(\widehat{d}_{\back}, \widehat{\theta}_\back)$.
\end{algorithmic}\end{minipage}
}
\end{algorithm}

\subsubsection{Step 3 (Clustering)}
We now identify the boundaries of the clusters in the delay and angle domain. 

\paragraph{Delay Domain}
For $i\in\{0,\back\}$, let $\bar{s}_i(\tau)$ denote the the angle-wise aggregated $\bar{S}^\JRAC_i(\tau,\theta)$, such that 
\begin{align} \label{eq:delay-aggregation}
    \bar{s}_i(\tau) & = \sum_{\ell = 1}^{|\bTheta|} \bar{S}_i^\JRAC(\tau,\theta_\ell),\; \forall \tau \in\calT_i, i\in\{0,\back\}.%\\
\end{align}
We then determine the non-zero points $t_{i,j}$ of the range-clusters and the boundary indicators $b_{i,j}$ in the delay domain as 
\begin{align}
    t_{i,j} & = \begin{cases}
        0, j = 0 \text{ or }  j = |\calT_i|+1,\\
        1, \bar{s}_i(\tau_j) \neq 0,\\
        0, \bar{s}_i(\tau_j) = 0,
    \end{cases}, j \in \{0,\ldots,|\calT_i|+1\}, \nonumber\\
    b_{i,j} & = t_{i,j+1} - t_{i,j}, j \in \{0,\ldots,|\calT_i|\}.
\end{align}
Note that for every $b_{i,\alpha} = 1,$ there is an associated $\beta > \alpha$, such that $b_{i,\beta} = -1$ and $b_{i,j''} = 0, \forall \alpha < j'' < \beta.$ This indicates the existence of a start and end boundaries of all delay-clusters. 

Let $\mbD_i$ define the set of the tuples $(\alpha_m,\beta_m)$ containing the index of the start boundary $\alpha_m$ and the end boundary $\beta_m$ of $m$-th delay-cluster. We define $m$-th delay-cluster as %by $\calD_{i,m}$ the $m$-th cluster defined as
\begin{equation}\label{eq:delay-cluster}
    \calD_{i,m} = \{j'' | \alpha_m < j'' < \beta_m, \exists (\alpha_m,\beta_m) \in \mbD_i\}. 
\end{equation}
Note that, $\cup_{m} \calD_{i,m} = \{j | t_{i,j} = 1\}$ represents all indices $j$ with non-zero values at $\bar{s}_i(\tau_j)$.

\paragraph{Angle Domain} Similarly, we denote the delay-wise aggregated $\bar{s}_i(\tau,\theta)$ by $\bar{s}_i(\theta)$, and determine the $n$-th angle-cluster $\calA_{i,n}$ using \eqref{eq:delay-aggregation}-\eqref{eq:delay-cluster}.

\subsubsection{Step 4 - Estimation}
Let the $k$-th cluster in the heatmap be defined using the $m$-th delay-cluster and $n$-th angle cluster as $\calC_{i,k} = \calD_{i,m} \times \calA_{i,n}$. The algorithm searches for the local maxima within each cluster as 
\begin{equation}
(\widehat{\tau}_{i,k},\widehat{\theta}_{i,k}) = \underset{(j,\ell)\in\calC_{i,k}}{\text{arg max}} \bar{S}_i^\JRAC(\tau_j,\theta_\ell). 
\end{equation}
Then, the cluster associated to the LoS path corresponds to the one with minimum delay. Therefore, 
\begin{align}
    k_i^* & = \underset{k}{\text{arg min}} \; \widehat{\tau}_{i,k},\\
    \widehat{\tau}_{0,0} & = \tau_{0,k_0^*}, \widehat{\tau}_{\back,0} = \tau_{\back,k_\back^*}, \widehat{\theta}_{\back, 0} = \theta_{\back,k_\back^*}.
\end{align}

Finally, following \eqref{eq::FFT-range-est}, 
\begin{equation}
    \widehat{d}_\back = \frac{c}{B} (\tau_{0,0} + \widehat{\tau}_{\back,0} - \widehat{\tau}_{0,0}), \quad\text{and} \quad \widehat{\theta}_\back = \widehat{\theta}_{\back,0},
\end{equation}
we get $(\widehat{d}_\back,\widehat{\theta}_\back)$, the resultant bistatic range-angle pair associated to the backscatter path.

The computational complexity of the JRAC algorithm is dominated by two steps: First, the MUSIC algorithm used in Step 1 for truncating the delay domain has complexity $\mathcal{O}(G_{\tau} N_\subc^{3})$.
Second, the computation of determining the heat map in~\eqref{eq::jointest1} has complexity $\mathcal{O}(G_{\tau} G_{\theta} N_\subc N_\ant)$. The Step 3 and 4 are one-dimensional searches with the complexity $\mathcal{O}(G_\tau+G_\theta)$. Therefore, the overall computational complexity of the JRAC algorithm is $\mathcal{O}(G_{\tau} N_\subc^{3} + G_{\tau} G_{\theta} N_\subc N_\ant)$.
\vspace{-0.15in}
\section{Positioning}\label{sec:positioning}
%------------------------------------
\begin{figure} 
\centering
\includegraphics[width=0.5\columnwidth]{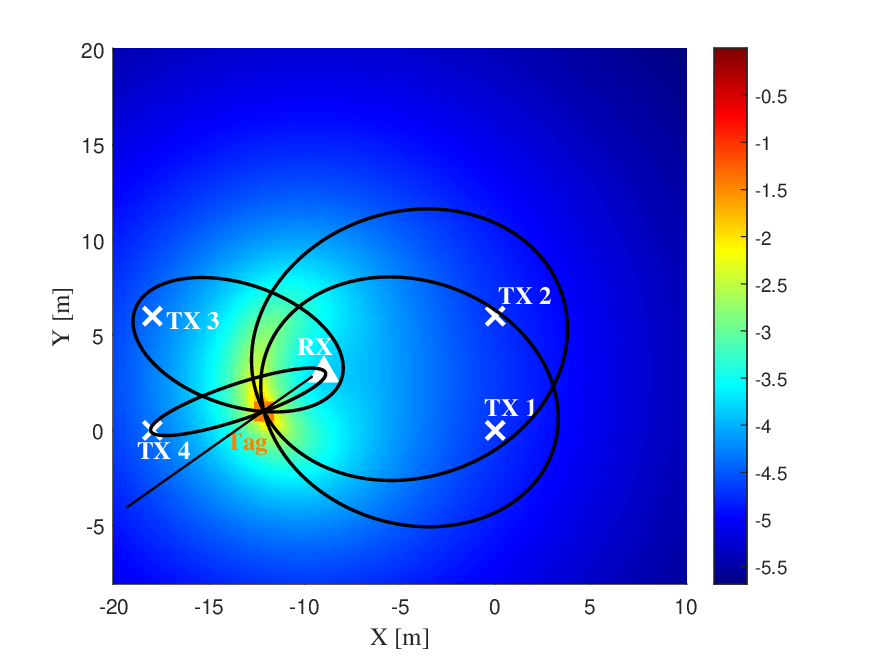}
\caption{Elliptical positioning scenario with bistatic range measurements from four TXs and one RX. The tag’s location is determined as the intersection of the four ellipses, each defined by a TX–RX pair as its foci and the hyperplane from the estimated tag AoA.}
\label{fig:ellipses}
\end{figure}
%------------------------------------
In this section, we propose two real-time multi-static positioning algorithms. As shown in Fig.~\ref{fig:ellipses}, the multi-static positioning scenario with range-angle measurements reduces to finding the intersection of the ellipses, each defined by a TX–RX pair as its foci, along with the hyperplane determined by the estimated tag AoA.  In the most general scenario, only a subset of receivers are equipped with antenna arrays that enable AoA estimation. Therefore, our subsequent formulation allows for range and angle measurements from different receivers to be exploited separately and does not assume that both range-angle estimates come from the same receiver. 

Let $\widehat{d}_{n}$ denote the bistatic range estimate of $n$-th measurement from the $i_n$-th TX and $j_n$-th RX. Denoting the noise in the range measurements as $\eta_{\text{r},n}$, we can write
\begin{equation}\label{eq:BR}
\widehat{d}_{n} = \| \mathbf{p}-\mathbf{p}_{{\TX,i_n}} \| +\|\mathbf{p}-\mathbf{p}_{{\RX,j_n}}  \| + \eta_{\text{r},n}.
\end{equation}
We assume $M_\text{r}$ number of range estimates for the positioning. 

Let $\widehat{\theta}_m,\widehat{\phi}_m$ denote the $m$-th azimuthal and elevation AoA measurements at the $j_m$-th RX. Therefore, the unit vector from the RX to the tag can be defined as
\begin{equation}\label{eq:angle_vec}
    \widehat{\mbv}_{m}=\begin{cases}
        [\cos{\widehat{\theta}_{m}},\sin{\widehat{\theta}_{m}}]^{\T}, & D= 2\\
        [\sin{\widehat{\phi}_m} \cos{\widehat{\theta}_{m}}, \sin{\widehat{\phi}_m}\sin{\widehat{\theta}_{m}}, \cos{\widehat{\phi}_m}]^{\T}, & D=3.
    \end{cases}
\end{equation}
% Given the ground truth unit vector $\mbu_m$ from \eqref{eq:GT-unitvector}, we define the noise in the $m$-th angle estimate as 
% \begin{equation}\label{eq:angle_error}
%     \eta_{\angle,m} = 1 - \widehat{\mbu}_m^\T \mbu_m.
% \end{equation}
We assume $M_{\angle}$ number of angle estimates for the positioning. 
Note that the AoA $\widehat{\theta}_{\back}$ in Section~\ref{sec:joint}, which is the estimated angle in the antenna frame, should be converted to angles defined in a global reference frame, as depicted in  Fig.~\ref{fig:usrp_map}. After AoA estimation and before positioning, a transformation matrix should be  applied on $\widehat{\mbv}_m$ to perform  conversion from the antenna frame  to global frame. We denote the  transformation matrix  by $\bOmega$ and the unit directional vector in  global frame will be $\widehat{\mbu}_m=\bOmega\widehat{\mbv}_m$.

\subsection{ML-based Positioning in  Multi-static BN using Range and angle Measurements}
We first introduce an ML-based positioning algorithm for multi-static BNs inspired from \cite{geng2021experimental}. 
\subsubsection{ML Formulation}
As in~\eqref{eq:BR}, we assume that $n$-th range measurement is corrupted by zero-mean Gaussian noise with variance $\sigma_{n}$, i.e., % \begin{align}
$\eta_{\text{r},n} \sim \mathcal{N}\left(0, \sigma_{n}\right).$
% \end{align}
Therefore, the log likelihood function of the tag position given the $n$-th range estimate, after ignoring additive constants \textit{w.r.t.} $\mbp$, is given by~\cite{kuntal2022ml}
\begin{align}
&\ell^{(\text{r})}_{n}(\mbp)=
%\log \left( \mathbb{P}\left[\widehat{d}_{n}\mid \mbp\right]\right)\\
%&=
-\frac{\left(\widehat{d}_{n}-\left\|\mathbf{p}_{{\TX,i_n}}-\mbp\right\|-\left\|\mathbf{p}{_{\RX,j_n}}-\mbp\right\|\right)^2}{2 \sigma_{n}^2}. 
\end{align}
Similarly, we assume that $m$-th angle measurement $\widehat{\theta}_m$ is corrupted such that the estimated directional estimate $\widehat{\mbu}_m$ follows von Mises-Fisher distribution with the mean direction $\mbu_m$ and the concentration parameter $\kappa_m$. Hence, \begin{align}
\widehat{\mbu}_{m} & \sim \text{VMF}(\mathbf{u}_m, \kappa_m), \quad \text{and}\\
\mathbb{P}(\widehat{\mbu}_m | \mathbf{u}_m, \kappa_m) & = \frac{\kappa_m}{4 \pi  \textrm{sinh}(\kappa_m)}\textrm{exp}(\kappa_m \widehat{\mbu}_m^{\T}\mbu_m).%\\ 
% \mathbb{P}(\widehat{\mbu}_m | \mathbf{u}_m, \kappa_m) & = \frac{\kappa_m}{4 \pi  \textrm{sinh}(\kappa_m)}\textrm{exp}(-\kappa_m     \eta_{\angle,m} ). 
\end{align}
Therefore, the log likelihood function of the tag position given the angle estimate $\widehat{\theta}_m$, after ignoring additive constants \textit{w.r.t.} $\mbp$, is given by 
\begin{align}
\ell^{(\angle)}_{m}(\mbp)&= \log \left(\mathbb{P}\left[\widehat{\mbu}_m\mid \mbp\right]\right)= \kappa_m \widehat{\mbu}_m^{\T}\frac{\mbp-\mbp_{{\RX,j_m}}}{||\mbp-\mbp_{{\RX,j_m}}||},
\end{align}
where we replace $\mbu_m$ by $(\mbp-\mbp_{{\RX,j_m}})/||\mbp-\mbp_{{\RX,j_m}}||$.
%\textcolor{blue}{Note that both $\hat{\mbu}_m$ and $\mbu_m$ are calculated with respect to the broadside direction of the  antenna array in the receiver.}
%{\color{red} IMPORTANT!}
% \ate{When evaluating the inner product in the VMF log-likelihood function, both vectors need to be in the same reference frame. Please note that the estimated direction $\widehat{\mbu}_m$ lies in the antenna frame, whereas the other unit vector $\mbu_m$ pointing to the tentative position $\mbp$ is in the World frame! You are missing the orientation matrix that transports the World vector into antenna frame. Please check my slides. Ort is the estimated direction $\widehat{\mbu}_m$ also in the World frame? Eq.~\eqref{eq:angle_vec} say it is in antenna frame. In any case angular estimated are obtained in the antenna frame, and need to be translated. I hope the implementation is correct.}

\begin{remark}
We adopt a directional statistics approach because the natural parameter space of angles is not an Euclidean space, but a sphere. Therefore, a distribution on the angles must account for the implicit periodicity of angles and must have a natural support over the unit sphere.  
The von Mises-Fisher distribution is the analog of the circular bi-variate normal distribution from the Euclidean space to the unit sphere. In fact, vMF distribution is equivalent to a bi-variate Gaussian distribution constrained onto a unit sphere in a $D$-dimensional space. The mean of such vMF distribution is the unit vector of the mean of the associated Gaussian distribution, and the reliability metric $\kappa_m$ is associated to the variance of the Gaussian distribution. Refer to \cite{directionalStat,abrudan2016vmf} for more details.
% Angles are periodic in their nature, and therefore, the natural support of the corresponding probability density functions should be the
% unit sphere. VMF measures dispersion from the mean directions, and the expression is tractable compared to warped normal distribution. Furthermore, VMF distribution models the    
\end{remark}
% % Errors in
% % estimating the directions of arrival are modeled by using the
% % von Mises-Fisher distribution
% % %\cite{geng2021experimental}, 
% % which is the correspondent of
% % the 2-D normal distribution to the two-dimensional unit sphere
% % \begin{equation}
% %     \operatorname{VMF}(\mathbf{u} \mid \boldsymbol{\mu}, \kappa)=\frac{\kappa}{4 \pi\textrm{sinh}(\kappa)}\exp \left(\kappa \boldsymbol{\mu}^{\mathrm{T}} \mathbf{u}\right),
% % \end{equation}
% % where $\bmu$ is the mean direction, and $\kappa$ is the concentration parameter. 
% The noisy directional
% estimates $\widehat{\mbu}_m$  are assumed to have a Von
% Mises-Fisher distribution with the mean direction $\mbu$ and concentration parameter $\kappa_m$ whose value reflects the
% reliability of the estimate. The noisy directional estimates follow the  distribution
% \begin{align}
% \mbu_{m} \sim \operatorname{VMF}(\mathbf{u}_m\mid \widehat{\mbu}_m, \kappa_m)= \frac{\kappa_m}{4 \pi  \textrm{sinh}(\kappa_m)}\textrm{exp}(\kappa_m \widehat{\mbu}_m^{\T}\mbu_m). 
% \end{align}
% where  $\widehat{\mbu}_m=[\cos{\widehat{\theta}_m},\sin{\widehat{\theta}_m}]^{\T}$. 
The log likelihood function $\mathcal{L}(\mbp)$ of the tag position given $M_\text{r}$-range and $M_{\angle}$-angle measurements is given by  \par \noindent \small
\begin{align}
&\calL(\mbp)= \sum_{n=1}^{M_\text{r}} \ell^{(\text{r})}_n(\mbp) + \sum_{m=1}^{M_{\angle}} \ell^{(\angle)}_m(\mbp),\nonumber\\
&= \sum_{n=1}^{M_\text{r}} -\frac{\left(\widehat{d}_{n}-\left\|\mathbf{p}_{\TX,i_n}-\mbp\right\|-\left\|\mathbf{p}_{\RX,j_n}-\mbp\right\|\right)^2}{2 \sigma_{n}^2}\nonumber\\
&+ \sum_{m=1}^{M_{\angle}}\kappa_m \widehat{\mbu}_m^{\T}\frac{\mbp-\mbp_{\RX,j_m}}{||\mbp-\mbp_{\RX,j_m}||}.
\end{align}
The tag's position can be estimated by maximizing the joint log likelihood function, i.e.,
\begin{align}\label{eq::ML-formulation}
\widehat{\mbp}_{\text{ML}}= \underset{\mbp\in \mathbb{R}^D}{\text{argmax}} \; \calL(\mbp).
\end{align} \normalsize
Since the likelihood function is not a convex function, a brute force approach using the grid search can be employed. This approach includes (1) quantizing the region of interest into grid points with resolution $\varepsilon_\text{ML}$, (2) evaluating the log likelihood function on each point in the grid, and (3) selecting the grid point that maximizes the the likelihood as a location estimate. While the grid search approach is guaranteed to provide a global optimal solution (up to the grid resolution), its computational cost increases inversely with $\varepsilon_\text{ML}$ as $\calO(D(M_\text{r} + M_\angle) \varepsilon^{-D}_\text{ML})$, where $\calO(D(M_\text{r} + M_\angle))$ is the cost of evaluating the log-likelihood function. 

For a practical deployment, we are interested in low complexity algorithms for real-time operations. Therefore, we introduce a gradient ascent-based heuristic algorithm to efficiently find the solution of \eqref{eq::ML-formulation}.

\subsubsection{ML Gradient Ascent with Line Search}

To perform gradient ascent, we first obtain the gradients of $\calL(\mbp)$ as \par \noindent \small
\begin{align}
&\nabla \mathcal{L}\left(\mathbf{p}\right)  \triangleq \frac{d}{d\mathbf{p}} \mathcal{L}(\mathbf{p})\\
& =-\frac{1}{2} \sum_{n=1}^{M_\text{r}} \frac{1}{\sigma_{n}^2}\Bigg[\left(\widehat{d}_{n}-\left\|\mathbf{p}_{\TX,i_n}-\mbp\right\|-\left\|\mathbf{p}_{\RX,j_n}-\mbp\right\|\right)\nonumber\\
&\hspace{0.9in}\left(\frac{\mathbf{p}_{\TX,i_n}-\mbp}{\left\|\mathbf{p}_{\TX,i_n}-\mbp\right\|}+\frac{\mathbf{p}_{\RX,j_n}-\mbp}{\left\|\mathbf{p}_{\RX,j_n}-\mbp\right\|}\right) \Bigg ] \\
&+ \sum_{m=1}^{M_{\angle}} \kappa_m \left(\frac{\mathbf{I}}{\|\mbp-\mathbf{p}_{\RX,j_m}\|} \right.\nonumber\\
&\hspace{0.6in}\left.-\frac{(\mbp-\mathbf{p}_{\RX,j_m})(\mbp-\mathbf{p}_{\RX,j_m})^{\T}}{\|\mbp-\mathbf{p}_{\RX,j_m}\|^3}\right) \widehat{\mbu}_m. 
\end{align} \normalsize
% since 
% \begin{align}
% &\nabla_{\mbp}  \left(\widehat{\mbu}_m^{\T}\frac{\mbp-\mbp_{_{\RX,j_m}}}{||\mbp-\mbp_{_{\RX,j_m}}||}\right) \\
% &=\left(\frac{\mathbf{I}}{\|\mbp-\mathbf{p}_{_{\RX,j_m}}\|_2}-\frac{(\mbp-\mathbf{p}_{_{\RX,j_m}})(\mbp-\mathbf{p}_{_{\RX,j_m}})^{\T}}{\|\mbp-\mathbf{p}_{_{\RX,j_m}}\|_2^3}\right) \widehat{\mbu}_m. \nonumber
% \end{align}

%The step $k$ of  gradient ascent will be 
%\begin{align}
%\widehat{\mbp}_{_{\text{ML}}}(k+1)= %\widehat{\mbp}_{_{\text{ML}}}(k)+\mu_k \nabla \calL(\mbp_k)
%\end{align}

Since the log-likelihood function is not a convex function, the gradient ascent algorithm is not guaranteed to provide the optimal solution. 
Therefore, we introduce two features in the gradient ascent to improve the possibility of reaching to the global maximum.

\noindent \textbf{(a) Multiple initializations:} The  gradient ascent iterations are initialized at the center, two vertices and two co-vertices of the ellipse associated to each TX-RX pair (see Fig.\ref{fig:ellipses}). This results in $5M_\text{r}$ initialization points denoted by $\widehat{\mbp}_{\text{ML},i}(0)$, $i=1,\ldots, 5M_\text{r}$. An independent instance of the gradient ascent algorithm is executed for each initial point.

\noindent \textbf{(b) Multiple step-sizes in each gradient ascent iteration}: For each instance of the gradient ascent, the estimate is updated at each iterations by selecting a step-size from a fixed set $\calS$ which maximizes the log-likelihood function. Specifically, 
\begin{equation}
\widehat{\mbp}_{\text{ML},i}(k+1) = \widehat{\mbp}_{\text{ML},i}(k)+\mu^*_k \nabla \calL(\widehat{\mbp}_{\text{ML},i}(k)),
\end{equation}
\begin{equation}
\text{where, } \mu^*_k = \underset{\mu_k \in \calS}{\text{arg max}} \; \; \mathcal{L}(\widehat{\mbp}_{\text{ML},i}(k)+\mu_k \nabla \calL(\widehat{\mbp}_{\text{ML},i}(k))). 
\end{equation}
Dynamically selecting the step-size in each iteration of  gradient ascent is also called gradient ascent with line search. At the end of each instance of the gradient ascent, the position that maximizes the likelihood is selected as the location estimate. As for the stopping criterion, we either  use $\varepsilon_{\text{ML}}$ as a predefined position accuracy threshold or a maximum number of iterations $K_{\text{ML}}$.
We summarize this algorithm, subsequently called \textit{Gradient Ascent with Line Search}, in Algorithm~\ref{alg::alg-gd}. 

%We emphasize here that the gradient ascent requires $\calO(1/\varepsilon_\text{ML})$ iterations to converge.

Given $5M_\text{r}$ initializations, $|\calS|$ evaluations of the log-likelihood functions in each iteration, and $\calO(D(M_\text{r}+M_\angle))$ complexity for the evaluation of the log-likelihood function, the total computational cost of the proposed ML-based gradient ascent with line search is given by $\calO(D(M_\text{r} + M_\angle)M_\text{r} |\calS| K_\text{ML})$.
%$\calO(D(M_\text{r} + M_\angle)M_\text{r} |\calS| \varepsilon_\text{ML}^{-1})$.

Note that, compared to ML-based grid search, the gradient search reduces the complexity by avoiding the exponential dependence on the number of dimensions $D$. This reduction in the complexity comes as a cost of the possibility of a sub-optimal solution and an additional linear dependence on $M_\text{r}$.

\begin{algorithm}[tb]
\caption{ML-based Gradient Ascent with Line Search}
\scalebox{0.8}{%
  \begin{minipage}{0.95\linewidth}
\begin{algorithmic}[1]
\Statex\textbf{Input} $\{\widehat{d}_{n}\}_{n=1}^{M_{\text{r}}}$, $\{\widehat{\mbu}_m\}_{m=1}^{M_{\angle}}$, $\calS$, $K$
\State \textbf{Initializations} $\{\widehat{\mbp}_{\text{ML},i}(0)\}_{i=1}^{5M_{\text{r}}}$%, set $k\gets 0$
\For{$i = 1$ to $5M_{\text{r}}$}
  \State $k \gets 0$
  \Repeat
    \State $\mu^*_k \gets \underset{\mu_k \in \calS}{\text{arg max}} \; \; \mathcal{L}(\widehat{\mbp}_{\text{ML},i}(k)+\mu_k \nabla \calL(\widehat{\mbp}_{\text{ML},i}(k)))$
    \State $\widehat{\mbp}_{\text{ML},i}(k+1) \gets \widehat{\mbp}_{\text{ML},i}(k)+\mu^*_k \nabla \calL(\widehat{\mbp}_{\text{ML},i}(k))$
    \State $k \gets k+1$
  \Until{$\|\widehat{\mathbf{p}}_{\text{ML},i}(k) - \widehat{\mathbf{p}}_{\text{ML},i}(k-1)\| \le \varepsilon_\text{ML}$ or $k=K_{\text{ML}}$}
\EndFor
\State $i^*\gets \underset{i\in\{1,\ldots,5M_\text{r}\}}{\text{arg max}} \; \; \calL(\widehat{\mbp}_{\text{ML},i}(k))$
\Statex \textbf{Output} $\widehat{\mbp}_{\text{ML}}\gets \widehat{\mbp}_{\text{ML},i^*}(k)$
\end{algorithmic}\label{alg::alg-gd}
\end{minipage}
}
\end{algorithm}\vspace{-0.2in}
\subsection{Iterative Re-weighted Least Squares (IRLS)}\label{sec::irls}
In this section, we propose an iterative positioning approach called \textit{IRLS}. While several prior works consider the IRLS algorithm~\cite{chan1994simple,amiri2017asymptotically,abrudan2024positioning}, they primarily focus on either range- or angle-only position estimation. We instead introduce the IRLS method that uses both range and angle estimates for multi-static positioning. 
% Further, our approach to IRLS can be extended to use range-angle estimates from multiple estimation algorithms, implicitly learn the quality of their estimates with respect to other estimates, and finally provide a robust positioning estimate.
Accordingly, we first formulate a pseudo-linear problem for multi-static positioning using range and angle measurements. We then introduce iterative weight updates, and finally, summarize the algorithm and discuss its computational complexity. 
\subsubsection{Pseudo-linear Formulation~\cite{chan1994simple}}
 % $\mathbf{p}_{{\TX,i_n}}$ and $\mathbf{p}_{{\RX,j_n}} $  denote  the position of the Tx and Rx that produced the  $\widehat{d}_n$ measurement. 

% \ate{Justifications to ignore the noise?}\kp{I think this is just a pre-cursor for pseudo-linear formulation. We do not claim that this assumption is indeed true. We are handling the noise through two steps described in (2) and (3) of this section.}
Assuming the noise $\eta_{\text{r},n} \ll \widehat{d}_n$, 
% \ate{What is $\eta_{\text{r},n}$ and why does it need to be less than 1? Is it the cosine similarity you removed? If so, please reformulate that equation holds in the LS sense, or similar.}
by re-arranging and squaring \eqref{eq:BR}, we get  \par \noindent \small
\begin{align}\label{eq:wls-ranges}
&\frac{1}{\brng[n]} \begin{bmatrix}
2 (\mathbf{p}_{{\RX,j_n}} - \mathbf{p}_{{\TX,i_n}})^{\top} & 2\widehat{d}_{n} \\
\end{bmatrix} \begin{bmatrix}
\mathbf{p}  \\
\| \mathbf{p} - \mathbf{p}_{{\RX,j_n}} \|
\end{bmatrix} = 1.
\end{align}
where \begin{equation}
    \brng[n] =\widehat{d}_{n}^2 -\| \mathbf{p}_{\TX,i_n} - \mathbf{p}_{\RX,j_n} \|^2+2 (\mathbf{p}_{\RX,j_n} - \mathbf{p}_{\TX,i_n})^{\top}\mathbf{p}_{\RX,j_n}.
\end{equation}
% where  $j_n \in \{1,\ldots,N_{{\mathrm{RX}}}\}$ and $i_n \in \{1,\ldots,N_{{\mathrm{TX}}}\}$. 
% \subsubsection{Positioning with  AoA  measurements}
%with respect to the $k$-th transmitter
Collecting \eqref{eq:wls-ranges} for all $M_{\text{r}}$ range measurements, we define
\begin{align}
\Arng &= [\Arng[1]^{\T}, \Arng[2]^{\T}, \ldots, \Arng[M_{\text{r}}]^{\T}]^{\T} \in \mathbb{R}^{M_{\text{r}} \times (D+N_\RX)}, \\
\Arng[n] &=
\frac{1}{\brng[n]} [2 (\mathbf{p}_{\RX,j_n} - \mathbf{p}_{\TX,i_n})^{\T}, \mathbf{0}_{j_n-1}^{\T},   2\widehat{d}_{n} , \mathbf{0}^{\top}_{N_\RX-j_n}].
\end{align} \normalsize
Similarly, assuming low angle measurement errors, we have 
\begin{equation}\label{eq:wls-angles}
\begin{bmatrix}
\mbI_{D} \; -\widehat{\mbu}_{m}
\end{bmatrix} \begin{bmatrix}
\mathbf{p} \\
\| \mathbf{p} - \mathbf{p}_{\RX,j_m} \|
\end{bmatrix}=\mathbf{p}_{\RX,j_m} \triangleq \bang[m].
\end{equation}
%Note that for   elevation angle  measurement is  available    the estimated direction is redefined as $\widehat{\mbu}_{m}=[\cos{\widehat{\theta}_{m}},\sin{\widehat{\theta}_{m}},\sin{\phi_m}]^{\T}$
% \subsubsection{Positioning with joint range and angle  measurements}
Collecting \eqref{eq:wls-angles} for all $M_\angle$ angle measurements, we define \par \noindent \small
\begin{align}
\Aang &= [\Aang[1]^{\T}, \ldots, \Aang[M_{\angle}]^{\T}]^{\T} \in \mathbb{R}^{D M_{\angle} \times (D+N_\RX)}, \\
\Aang[m] &=[\mbI_D, \bzero_{D\times (j_m-1)},
-\widehat{\mbu}_{m}, \mathbf{0}^{\top}_{D\times N_\RX-j_m}], \\
\bang &=[\bang[1]^{\T},\ldots,\bang[M_{\angle}]^{\T}]^{\T} \in \mathbb{R}^{D M_{\angle} \times 1}.
\end{align}
Finally, let 
\begin{align}
\bUpsilon&= \begin{bmatrix} \mathbf{p}^{\T},\|\mbp-\mbp_{\RX,1}\|, \ldots, \|\mbp-\mbp_{\RX,N_\RX}\| \end{bmatrix}^{\T},\label{eq:ls-theta}\\% \in \mathbb{R}^{D+N_\RX},  \\
\Aall &= \begin{bmatrix}\Arng^\T , \Aang^\T
\end{bmatrix}^\T \in \mathbb{R}^{(M_{\text{r}}+DM_{\angle}) \times (D+N_\RX)}, \label{eq:ls-A}\\
\ball &=
\begin{bmatrix}
\mathbf{1}_{M_\text{r}}^\T,  \bang^\T
\end{bmatrix}^\T    \in \mathbb{R}^{(M_{\text{r}}+DM_{\angle})\times 1}. \label{eq:ls-B}
\end{align}
From \eqref{eq:wls-ranges} and \eqref{eq:wls-angles}, \eqref{eq:ls-theta}-\eqref{eq:ls-B} can be collectively written as
\begin{align} \label{eq:ls}
\Aall \bUpsilon&=\ball.
\end{align} \normalsize
To estimate the tag position, we need to estimate $\bUpsilon$ using a least square solution given $\Aall$ and $\ball$. 
This provides a pseudo-linear\footnote{\eqref{eq:ls} is a pseudo-linear formulation because the parameters defined in $\bUpsilon$ are not independent variables, but have non-linear dependency between them.} formulation of the multi-static positioning problem, given the range and angle measurements.
% We emphasize here that \eqref{eq:ls} is indeed a pseudo-linear formulation. The variable $\bUpsilon$ does not contain independent variables. While this 
Furthermore, this formulation can also be extended to include the measurements from  multiple methods simply by stacking more rows in the matrix $\Aall$ and $\ball$. 

\begin{remark}
    In addition to using both range and angle measurements, our IRLS formulation differs from the prior work in another way. 
    All rows corresponding to the range measurements in \eqref{eq:ls} are scaled such that their respective elements in $\ball$ is 1. 
    From \eqref{eq:wls-ranges}, note that $\brng[n]$ depends on the locations of the TXs and RXs, and $\widehat{d}_n$. 
    While most prior work focused on simulations which resulted in comparable $\brng[n]$ across all measurements, our experience with the real-world prototype showed that, in practice, $\brng[n]$ varies drastically across measurements. 
    Hence, solving \eqref{eq:ls} without the scaling by $\brng[n]$ would give artificially high importance to the measurements with high $\brng[n]$. 
    This problem is further exacerbated if $n$-th range measurement is an outlier resulting in high $\widehat{d}_n^2$, which would add higher weight to an outlier measurement! 
    Therefore, we tackle this problem (highlighted by our experimental setup) by simply scaling \eqref{eq:wls-ranges} by $\brng[n]$ such that the  right hand side is one.
\end{remark}
%\ate{I still find this weird. I have used IRLS many times, I have never encountered such a situation.}

While a least squares solution   solves \eqref{eq:ls}, from a practical perspective, there are two key challenges: 
%\begin{enumerate}
%    \item

$\bullet$ \textit{Errors in the measurements:} \eqref{eq:ls} does not account for the possible errors in the range and angle measurements due to signal noise and multipaths in the propagation environment.
%\item

$\bullet$ \textit{Validity of estimated $\bUpsilon$:} The least square solution of \eqref{eq:ls} does not enforce the dependency within $\{\bUpsilon_i\}_{i=1}^{D+N_\RX}$. Specifically, the solution $\widehat\bUpsilon$ is only valid if 
    \begin{equation}\label{eq::constraints}
            \|[\widehat{\bUpsilon}]_{1:D}-\mbp_{\RX,j} \|= [\widehat{\bUpsilon}]_{D+j}, \forall j\in \{1,\ldots,N_\RX\}.
    \end{equation}
%\end{enumerate}
In the following, we introduce two methods to tackle these two challenges.  

\subsubsection{Iterative Re-Weighted Least Square}
The formulation in \eqref{eq:ls} assumes all measurements to have same importance. 
However, an error in one measurement may distort the estimate. Therefore, we introduce a measure of reliability for each measurement. 

We define a weight vector $\mbw \in \mathbb{R}^{M_{\text{r}}+D\mathbb{R}^{M_{\angle}}}$ as the weights assigned to each row of \eqref{eq:ls}. Then, with $\mbW=\text{Diag}\{ \mbw \}$, 
we solve
\begin{equation}\label{eq:wls}
\widehat{\bUpsilon} = \underset{\bUpsilon}{\text{min}} \quad (\ball - \Aall \bUpsilon)^{\T} \mathbf{W} (\ball - \Aall \bUpsilon),
\end{equation}
which results in
\begin{equation}\label{eq:wls-solution}
% \widehat{\bUpsilon}=\left(\Aall^{\T}\mbW \Aall\right)^{-1}\Aall^{\T}\mbW \ball.
\widehat{\bUpsilon}=\left(\sqrt{\mbW} \Aall\right)^{\dagger}\sqrt{\mbW} \ball.
\end{equation}
The position estimate is then given by
\begin{equation}\label{eq:IRLS-first}
\widehat{\mbp}_\text{IRLS}=[\widehat{\bUpsilon}]_{1:D}.
\end{equation}

To estimate the position $\widehat{\mbp}_{\text{IRLS}}$ robustly, our goal is to select the weights $\mbw_\text{r}$ and $\mbw_\angle$ such that the outliers are assigned lower weight. For that, we use an iterative approach as follows:\begin{enumerate}
\item Initialize the weight vector by $\mbw = \mathbf{1}_{M_{\text{r}}+D\mathbb{R}^{M_{\angle}}}$. 
\item Estimate $\widehat{\bUpsilon}$ using \eqref{eq:wls-solution}.
\item Estimate the residual errors $\mbe = \left|\Aall\bUpsilon - \ball\right|$. 
\item Define $\mbw$ such that $\mbw_i = (\mbe_i^2 + \varepsilon_{\text{IRLS}})^{-1}$. 
\item Repeat 2-4 for $K$ number of iterations,
\end{enumerate}
where $\varepsilon_{\text{IRLS}}$ prevents excessively large weights and ensures stability.
We emphasize here that while we set same initial weights for all measurements, in practice, there are other approaches using measurement metrics. For instance, the measurements from the tag packets with higher signal strength quality or lower bit error rate can be assigned higher initial weights. However, given that IRLS does not guarantee an optimal solution, it is not possible to definitively conclude the best choice for the initial weights.  

% In order to iteratively  update the assigned  weights, we define the residual
% errors for range estimates as 
% \begin{equation}
% e_{r,n}=|\widehat{d}_n-(\| \widehat{\mbp}_{_\text{IRLS}}-\mathbf{p}_{_{\TX,i_n}} \| +\|\widehat{\mbp}_{_\text{IRLS}}-\mathbf{p}_{_{\RX,j_n}}  \|)|   
% \end{equation}
% which represents the difference between the measured bistatic ranges and the
% actual bistatic ranges with respect to the estimate of the user's position $\widehat{\mbp}_{_{\text{IRLS}}}$.
% and the the residual
% errors for angle  estimates as 
% \begin{equation}
% e_{\angle,m}= \|\mbp -\mbp_{_{\RX,j_m}} - \widehat{\mbu}_m  \| \mathbf{p} - \mathbf{p}_{_{\RX,j_m}} \|   \|
% \end{equation}  
% The  above  residual errors may be interpreted as a measure
% of how much the range and angle measurements  agree with the
% current position estimate $\mbp_{_{\text{IRLS}}}$. After computing all residuals,
% the weights for the next iteration are are  updated  as
%  $\mbw_{r,n}=(e_{r,n}^2+\varepsilon_{\text{IRLS}})^{-1}$ and $\mbw_{\angle,m}=(e_{\angle,m}^2+\varepsilon_{\text{IRLS}})^{-1}$, where $\varepsilon_{\text{IRLS}}$ is a small number near machine precision. 
% The IRLS procedure is finished if a certain number of iterations is reached, or the difference between
% the current and the previous position estimate is below a pre-defined threshold. For initializing the weights, higher values are assigned to measurements with higher
% RSSI.
\subsubsection{Ensuring Validity of the Estimate}
The solution of~\eqref{eq:wls-solution} assumes that the elements of $\bUpsilon$ are independent. However, the estimated $\widehat{\bUpsilon}$ is valid only if \eqref{eq::constraints}  is satisfied. 
%\begin{equation}\label{eq::constraints_1}
%\|[\widehat{\bUpsilon}]_{1:D}-\mbp_{\RX,j} \|= [\widehat{\bUpsilon}]_{D+j}, \forall j\in\%{1,\ldots,N_\RX\}.
%\end{equation}
To ensure that this constraint is satisfied, we use a second stage least squares solution ~\cite{chan1994simple,amiri2017asymptotically} to project $\hat{\bUpsilon}$ on to feasible cone of \eqref{eq::constraints}. Let $\boldsymbol{\eta}_\text{IRLS}$ represent the error between the estimated $\widehat{\bUpsilon}$ and the true $\bUpsilon$ i.e. $\widehat{\bUpsilon}=\bUpsilon+\boldsymbol{\eta}_\text{IRLS}$. We define $\bUpsilon_2 = \mbp_2 \odot \mbp_2$ as the unknown of the second stage least squares and  correspondingly  $\mbp_2$ as  the tag position estimate after projection. Then, \par \noindent \small
\begin{align} \label{eq::irls2}
    \bUpsilon_2=  [\widehat{\bUpsilon}]_{1:D} \odot [\widehat{\bUpsilon}]_{1:D} - 2[\boldsymbol{\eta}_\text{IRLS}]_{1:D} \odot [\widehat{\bUpsilon}]_{1:D}. 
\end{align} \normalsize
Squaring~\eqref{eq::constraints} yields,
\begin{align} \label{eq::irls3}
\mathbf{1}^\T \bUpsilon_2 & - [\widehat{\bUpsilon}]_{D+j}^2 -2 \mbp_{\RX,j}^\T [\widehat{\bUpsilon}]_{1:D} + ||\mbp_{\RX,j}||^2& \nonumber \\
    &= -2[\widehat{\bUpsilon}]_{D+j}[\boldsymbol{\eta}_{\text{IRLS}}]_{D+j}- 2\mbp_{\RX,j}^\T[\boldsymbol{\eta}_{\text{IRLS}}]_{1:D},& 
\end{align} \normalsize
for $j=\{1,\ldots,N_{\RX}\}$. In~\eqref{eq::irls2}-\eqref{eq::irls3}, we  use  $\bUpsilon=\widehat{\bUpsilon}-\boldsymbol{\eta}_\text{IRLS}$ and elliminated the second-order error terms. Consequently, \eqref{eq::irls2}-\eqref{eq::irls3} are recast in matrix form as $\Aall_2 \bUpsilon_2 = \ball_2$, where
\par \noindent \small 
\begin{align}
%\bUpsilon_2 & = \mbp_2 \odot \mbp_2, \\
\Aall_2 &= [\mbI_D, \mathbf{1}_D, \ldots,\mathbf{1}_D]^\T \in \mathbb{R}^{(D+N_\RX)\times D}\label{eq:Aall_2}, \\
\ball_2 & = \widehat{\bUpsilon} \odot \widehat{\bUpsilon} + \begin{bmatrix}
    \mathbf{0}_D \\ 
    2 \mbp_{\RX,1}^\T [\widehat{\bUpsilon}]_{1:D} - ||\mbp_{\RX,1}||^2 \\
    2 \mbp_{\RX,2}^\T [\widehat{\bUpsilon}]_{1:D} - ||\mbp_{\RX,2}||^2 \\
    \vdots\\
    2 \mbp_{\RX,N_\RX}^\T [\widehat{\bUpsilon}]_{1:D} - ||\mbp_{\RX,N_\RX}||^2.
    \end{bmatrix} \label{eq:ball_2}
\end{align} \normalsize
Then, we find $\mbp_2$ such that \par \noindent \small 
\begin{equation}
    \widehat{\mbp}_2 = \underset{\mbp_2\in\mathbb{R}^D}{\text{arg min}}\; ||\Aall_2 \bUpsilon_2 - \ball_2||^2,
\end{equation} \normalsize
which results in the solution
\begin{equation}
 \widehat{\bUpsilon}_2=   \widehat{\mbp}_2 \odot \widehat{\mbp}_2 = {\Aall_2^\dagger \ball_2}.
\end{equation}
Therefore, the final position estimate from the IRLS algorithm can be given as \par \noindent \small
\begin{equation}\label{eq:IRLS-final}
    \widehat{\mbp}_{\text{IRLS}} = \text{sign}([\widehat{\bUpsilon}]_{1:D})\odot \sqrt{\Aall_2^\dagger \ball_2}.
\end{equation} \normalsize

Note that \eqref{eq:IRLS-final} improves upon the estimate \eqref{eq:IRLS-first} by fulfilling the constraint introduced in \eqref{eq::constraints}. 

\subsubsection{Algorithm Complexity}
%The algorithm for IRLS positioning using range and angle measurements is summarized in Algorithm~\ref{alg::alg-irls}. 
%---------------------------------------------
% \begin{algorithm}[tb]
% \caption{IRLS for joint   range and angle positioning }
% \scalebox{0.75}{%
%   \begin{minipage}{0.95\linewidth}
% \begin{algorithmic}[1]
% \Statex\textbf{Input} $\{\widehat{d}_{n}\}_{n=1}^{M_{\text{r}}}$, $\{\widehat{\mbu}_m\}_{m=1}^{M_{\angle}}$, $K$
% \State \textbf{Initialization:}  A diagonal matrix $\mbW^{(0)}$ with entries 1.
% \For{$i = 1$ to $K$}
%   % \State $\widehat{\bUpsilon}^{(i)} \gets\left(\Aall^{\T} \mbW^{(i)} \Aall\right)^{-1}\Aall^{\T}\mbW^{(i)} \ball$
%   \State $\widehat{\bUpsilon}^{(i)}\gets\left(\sqrt{\mbW^{(i)}} \Aall\right)^{\dagger}\sqrt{\mbW^{(i)}} \ball.$
%   \State $\mbe \gets \left|\Aall \widehat{\bUpsilon}^{(i)} - \ball\right|$
%   \State $\mbW^{(i+1)}_{n,n}\gets(\mbe_{n}^2+\varepsilon_{\text{IRLS}})^{-1}, \forall n\in\{1,\ldots,M_\text{r}+DM_\angle\}.$
%   \State $$\mbW^{(i+1)}_{n,n}\gets\frac{\mbW^{(i+1)}_{n,n}}{\sum_{n'}\mbW^{(i+1)}_{n',n'}}, \forall n\in\{1,\ldots,M_\text{r}+DM_\angle\}.$$
% \EndFor
% \State Use $\widehat{\bUpsilon}^{(K+1)}$ to obtain $\Aall_2$ and $\ball_2$ from \eqref{eq:Aall_2} and \eqref{eq:ball_2}.
% \Statex \textbf{Output} $\widehat{\mbp}_{\text{IRLS}} \gets \text{sign}\left([\widehat{\bUpsilon}]^{(K+1)}_{1:D}\right) \odot \sqrt{\Aall_2^\dagger \ball_2}.$
% \end{algorithmic}\label{alg::alg-irls} \end{minipage}
% }
% \end{algorithm}
%-----------------------------------------------
The computation for IRLS algorithm is dominated by the pseudo-inverse operation defined on line 3 of the algorithm. Given that the dimensions of $\Aall$ is $(M_\text{r} + DM_\angle)\times (D+N_\RX)$, the complexity of the pseudo-inverse operation is $\calO(D(M_\text{r} + DM_\angle)^2).$ Therefore, the total complexity of the IRLS algorithm is given by $\calO(KD(M_\text{r} + DM_\angle)^2)$.

\section{Numerical Results}
% \begin{table}[tb]
% \centering
% \caption{Real-time computational complexity of joint range/AoA estimation methods}
% \label{tab:realtime_complexity_joint_range_aoa}
% \begin{tabular}{c|c|c}
% \hline
% \textbf{Estimation Method} & \textbf{Computational Complexity} & \textbf{Average Run time(ms)} \\
% \hline
% SRAE & $\mathcal{O}()$ &  \\
% \hline
% JRAC & $\mathcal{O}()$ &  \\
% \hline
% 2D MUSIC& $\mathcal{O}(G N_\ant^3 N_\subc^3)$ &  \\
% \hline
% 2D FFT & $\mathcal{O}()$ &  \\
% \hline
% IR First & $\mathcal{O}()$ &\\
% \hline
% \end{tabular}
% \end{table}

In this section, we present the simulated performance of the proposed estimation and localization algorithms.
%, SRAE and JRAC, and the positioning algorithms, ML- and IRLS-based algorithms. 
\vspace{-0.1in}
\subsection{Simulation Setup}
Four TXs are located at the coordinates $(0,0)$, $(0,6)$, $(-18,6)$, and $(-18,0)$~m within the room. The RX, equipped with an antenna array with $N_\ant=4$ half-wavelength–spaced elements, is positioned at $(-9,3)$~m. 
25 tags are placed uniformly in the region $(x,y) \in [-18,0] \times [0,6]$~m. All TXs, RX and tags are assumed to have same height, making the entire deployment a 2D deployment.

Scatterers are uniformly distributed in the same region as the tag positions. We assume $L_1,L_2=3$ number of scatterers in the environment. Further, we generate 50 OFDM symbols, with 40 subcarriers uniformly spaced at 1 MHz, according to \eqref{eq::vec-h}. We further contaminate the OFDM symbols by Gaussian noise $\mbv^{(n)}\sim \calN(0,\sigma_v^2)$ such that the SNR$=\frac{\|\boldsymbol{\calH}_{\back,n}^{(n)}\|}{\sigma_v^2}$ is 5 dB.% for all symbols. 
% We generate synthetic CFR measurements for $N_\sym = 50$ symbols considering 40 subcarriers uniformly spaced at 1 MHz. % in \tabref{tab:sim_params}. 
% \begin{table}[tb]
% \centering
% \caption{Parameters for Synthetic Data Generation}
% \label{tab:sim_params}
% \begin{tabular}{l|l}
% \hline
% \textbf{Parameter} & \textbf{Value} \\ \hline
% Number of subcarriers  $N_\subc$& 40 \\ \hline
% OFDM subcarrier spacing $B/N_\subc$& 1 MHz \\ \hline
% Number of OFDM symbols  $N_{\sym}$ & 50 \\ \hline
% Number of  Scatterers  $L_1, L_2$ & 3 \\ \hline
% Number of  Antenna in ULA $N_\ant$  & 4 \\ \hline
% Signal to Noise Ratio  at the receiver  & 5 dB \\ \hline
% \end{tabular}
% \end{table}
\vspace{-0.1in}
\subsection{Range-Angle Estimation Performance}
We evaluate joint range–angle estimation methods from \secref{sec:joint}: 2D FFT, 2D MUSIC, SRAE with $T^\SRAE_{\textrm{min}}=0.5$, and JRAC with $T^\JRAC_{\textrm{min}}=0.2$ and $T^\JRAC_{\textrm{max}}=0.6$. We also add a range-only estimation using IR First~\cite{abrudan2025next} as a baseline to emulate a single-antenna RX. For all algorithms, we use same sets $\calT$ and $\bTheta$ with $G_\tau = 4096$, $G_\theta = 128$. This ensures that performance and runtime of proposed algorithms and IR First are fairly compared.

\begin{figure*}[tb]
\centering
\subfloat[CDF of range estimation error\label{fig:sim-range-error}]{\includegraphics[height=1.3in]{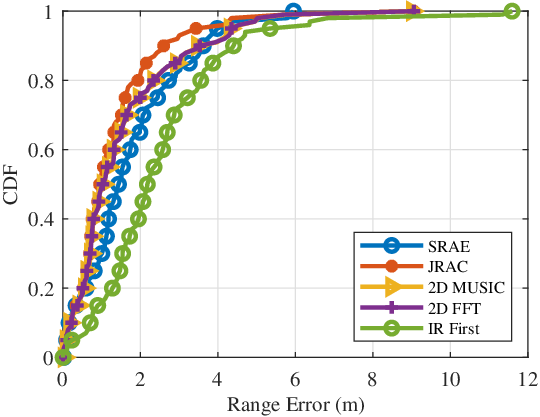}}\quad
\subfloat[CDF of angle estimation error\label{fig:sim-angle-error}]{\includegraphics[height=1.3in]{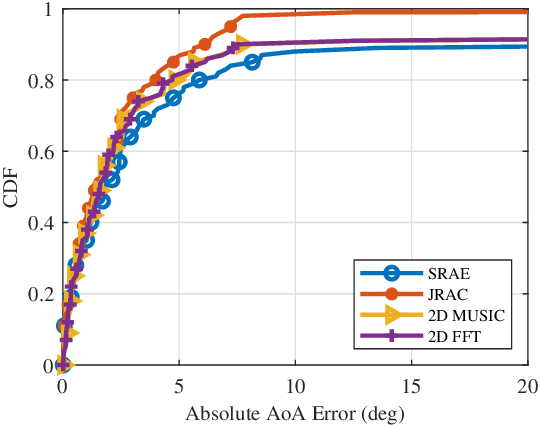}}\quad
\subfloat[CDF of localization error\label{fig:sim-loc-error}]{\includegraphics[height=1.3in]{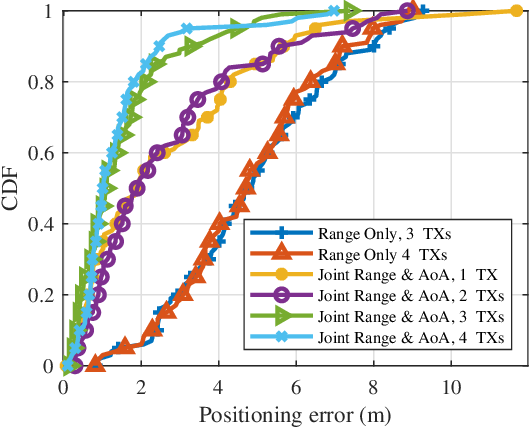}}
\caption{(a)-(b) Range and angle estimation errors using simulation data: Range-AoA estimation methods have higher ranging accuracy compared to the range-only method IR-First. Joint methods such as JRAC achieves higher ranging and angle estimation accuracy compared to the stage-wise method, SRAE. (c) Using AoA for localization with various number of TXs: We compare positioning accuracy using range only and joint range-angle measurements from the synthetic data. We consider SNR$=5$ dB, $L_1=L_2=3$ scatterers, 1D MUSIC for the range estimation and 2D MUSIC for joint range and angle estimation. Localization uses ML-based Gradient Ascent with Line Search.}
\label{fig:sim_results}
\end{figure*}

\figref{fig:sim_results} compares the performance of these five algorithms in terms of the absolute range estimation error $|\widehat{d}-d|$ and the absolute angle estimation error $|\widehat{\theta}-\theta|$. 
\figref{fig:sim-range-error} shows that the range-angle estimation algorithms outperforms the range-only baseline. This indicates the benefit of separating the multipath components along the spatial dimensions to improve the range estimates. 
Among the range-angle estimation algorithms, both \figref{fig:sim-range-error} and \ref{fig:sim-angle-error} shows (a) JRAC outperforming 2D FFT and 2D MUSIC, and (b) SRAE under-performing 2D FFT and 2D MUSIC. JRAC outperforms due to the clustering step which considers a multiple paths as a unit instead of evaluating individual paths as done by 2D FFT and 2D MUSIC. Furthermore, SRAE under-performs the joint methods due its stage-wise architecture. In theory, since delay and angle are coupled in multipath environments, joint algorithms can separate them more accurately.
%------------------------------------
\vspace{-0.1in}
\subsection{Positioning Performance}
The benefit of positioning based on  joint range and angle measurements from synthetically generated dataset, compared to range-only measurements, is illustrated in~\figref{fig:sim-loc-error}. In a two-dimensional positioning scenario, range-only localization requires a minimum of three ellipses corresponding to three TXs to estimate a tag position. In contrast, when the RX is equipped with an antenna array and the AoA estimates are available, the tag position can be determined using as few as one range and one angle measurements from a single TX. \figref{fig:sim-loc-error} further shows that even with a sufficient number of TXs, equipping the RX with a ULA to enable AoA estimation significantly improves positioning accuracy. Specifically, with three TXs, range-only measurements result in an average positioning error of approximately $6$m, which is reduced to about $2$m when AoA estimation at the RX is feasible.
\vspace{-0.1in}
% \begin{figure}[tb]
% \centering
% \includegraphics[width=.7\columnwidth]{Images/localization_error_SNR5_3sc_ch.eps}
% \caption{Benefit of AoA in terms of number of TXs. We compare positioning accuracy using range only and joint range-angle measurements from the synthetic data. We consider SNR$=5$ dB, $L_1=L_2=3$ scatterers, 1D MUSIC for the range estimation and 2D MUSIC for joint range and angle estimation. Localization uses ML-based Gradient Ascent with Line Search.}
% \label{fig:all}
% \end{figure}
\section{Experimental Results}
\label{sec:experimental_results}
%In this section, we evaluate 
%the range-angle estimation and positioning algorithms using a large-scale testbed.\vspace{-0.1in}
\begin{figure*}
    \centering
    \includegraphics[height=1in]{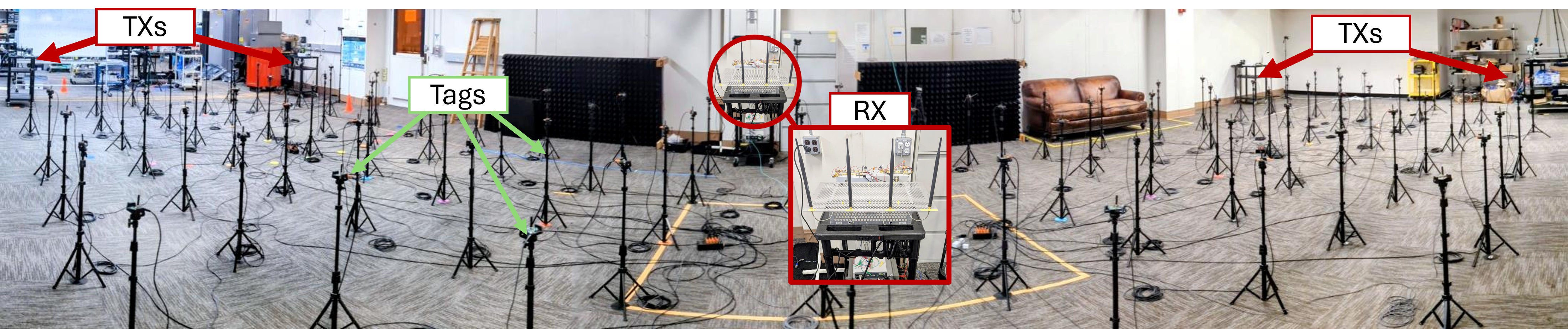}
    \caption{The experimental setup of a large-scale BN consisting of 4 TXs, 1 RX with 4-element antenna array, and 100 tags in an enclosed lab. The walls with metal surface and the ground introduce significant multipath in the channel.}
    \label{fig:testbed}
    \vspace{-0.2in}
\end{figure*}
\begin{figure} 
\centering
\includegraphics[width=0.95\columnwidth]{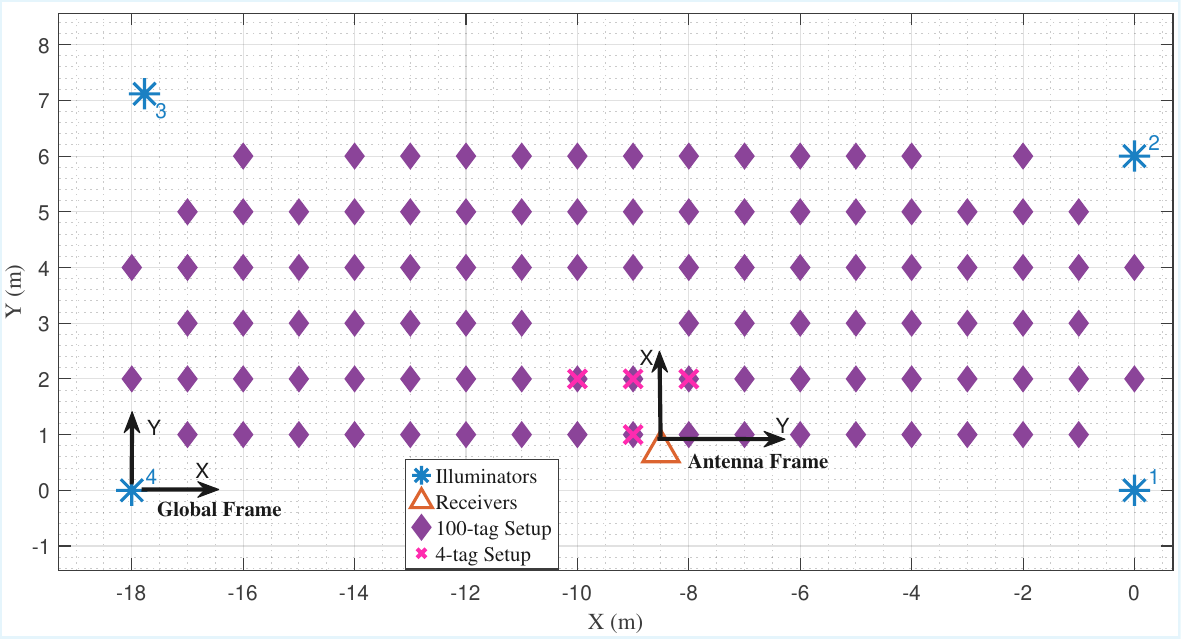}
\caption{Schematic of the experimental setup consisting 4 TXs, 1 RX with 4-element antenna array, and 100 tags. The positive-X direction in the antenna frame denotes the broadside direction of the receiver antenna array.}
\label{fig:usrp_map}
\end{figure}
% \begin{figure}
%     \centering
%     \includegraphics[width=.6\columnwidth,trim={.2cm 0cm 0cm 0.2cm},clip]{Results_USRP/ReceiverSetup.pdf}
%     \caption{Schematic of the RX using USRP X440: RF duplexers split the incident signal into two bannds associated to carrier and backscatter signals.}
%     \label{fig:usrp_schematic}
% \end{figure}
%\subsection{Experimental Setup}
We conduct the experimental evaluation on a custom-built wireless testbed deployed in an enclosed lab space (metal walls). %concrete floor). 
The testbed spans an area of 20 m × 7 m and includes four TXs, a single multi-channel RX, and 100 custom-designed semi-passive backscatter tags distributed across the area. The TXs and RX operate without any clock synchronization, reflecting a realistic multi-static architecture without clock distribution. The experimental setup is given in Fig.~\ref{fig:testbed}.

Each TX uses an ADRV9361-Z7035 chip connected to a dipole antenna and continuously broadcasts an OFDM waveform with 39.36 MHz bandwidth centered at 897.5~MHz. The OFDM waveform contains 41 subcarriers spaced at 960 kHz, following the design principles in \cite{abrudan2025next}.% \tabref{tab:usrp_params} summarizes the waveform parameters. 

% to provide sufficient bandwidth for high-resolution channel estimation and localization.

% \begin{table}[tb]
% \centering
% \caption{Waveform Parameters for Experimental Setup}
% \label{tab:usrp_params}
% \begin{tabular}{l|l}
% \hline
% \textbf{Parameter} & \textbf{Value} \\ \hline
% Number of subcarriers $N_\subc$& 41 \\ \hline
% OFDM subcarrier spacing & 960 kHz \\ \hline
% Number of OFDM symbols $N_{\sym}$ per tag packet & 98 \\ \hline
% Sampling rate & 61.44 MSps \\ \hline
% \end{tabular}
% \end{table}

The backscatter tags have the semi-passive architecture described in \cite{abrudan2025next}. Each tag transmits one 100-bit packet of duration 4.8~ms at a uniformly random time within every 300~ms window~\cite{patel2025scalability}. Tags modulate the incident OFDM carrier using a square-wave switching mechanism and apply a frequency shift of~45 MHz to separate the backscatter signal from the carrier, which prevents interference with the direct signal from the TX. 

The receiver is built around a USRP X440 with eight RF channels. A 4-element half-wavelength-spaced custom ULA feeds into 4 RF diplexers, which split each antenna feed into two bands containing: (a) carrier signal from the direct TX-RX path, and (b) backscatter signal from the TX-tag-RX path. 
This configuration produces eight input streams to the USRP. Since USRP X440 lacks internal amplification, we add external LNAs with 20 dB gain each, which boosts the SNR of weak backscatter signals. 
% \figref{fig:usrp_schematic} shows the schematic of the receiver architecture. 
The USRP samples all eight streams at an ADC rate of 61.44~MSps and sends them to the host computer for processing. I/Q processing follows the steps described in \cite{abrudan2025next}. The RX estimates CFRs for all streams, estimates the phase and timing references from the four carrier channels, and computes the range and angle estimates from the four backscatter channel estimates of each detected tag. Then, the estimated tag direction  in the antenna frame is transformed to the world frame as explained in~\eqref{eq:angle_vec}. For the topology illustrated in Fig.~\ref{fig:usrp_map}, the transformation matrix is
$
\boldsymbol{\Omega}
=
\left[
\begin{array}{cc}
0 & 1 \\
1 & 0
\end{array}
\right]
$.
Finally, the range and angle estimates for each tag using the carrier signals from all 4 TXs are used to estimate the position of the tags.
\vspace{-0.1in}
%\subsection{Scenarios}
We evaluate two scenarios as shown in \figref{fig:usrp_map}: \begin{itemize}
    \item \textbf{4-tag setup}: This setup presents an ideal condition with high SNR and few collisions. We keep four tags close to the RX on and capture 1.2 seconds of continuous samples.
    \item \textbf{100-tag setup}: This setup emulates a realistic deployment with low SNR and dense deployment. We keep all 100 tags on and capture 15 seconds of continuous samples to ensure sufficient observations for evaluation. 
\end{itemize}
In both scenarios, the TXs transmit continuously and the RX records raw I/Q samples for processing.
%In the subsequent sections, we present the accuracy of range-angle estimation and positioning using the proposed algorithms. \vspace{-0.1in}
%------------------------------------
%\subsection{Range-Angle Estimation Performance}\label{sec:usrp_est}
We evaluate four range–angle estimation algorithms described in \secref{sec:joint}: (a) 2D FFT, (b) 2D MUSIC, (c) SRAE with $T^\SRAE_{\textrm{min}}=0.5$, and (d) JRAC with $T^\JRAC_{\textrm{min}}=0.2$ and $T^\JRAC_{\textrm{max}}=0.6$. Similar to \figref{fig:sim-range-error}, we also add a range-only estimation using IR First~\cite{abrudan2025next} as a baseline to emulate a single-antenna RX setup. For all algorithms, we use same sets $\calT$ and $\bTheta$ with $G_\tau = 4096$ and $G_\theta = 128$. This ensures that  performance and runtime of proposed algorithms and IR First are fairly compared.

\begin{figure*}[tb]
    \centering
      % \begin{minipage}{.3\linewidth}
\subfloat[Range est. error, 4-tag setup\label{fig:usrp_range_est_4_tags}]{%
\includegraphics[height=1.3in]{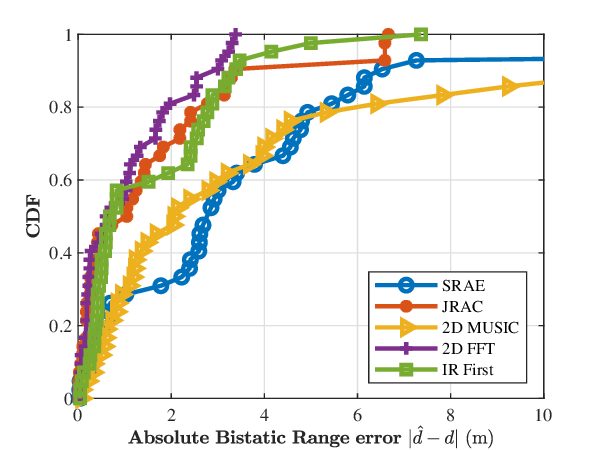}}\quad
\subfloat[Angle est. error, 4-tag setup\label{fig:usrp_ang_est_4_tags}]{%
\includegraphics[height=1.3in]{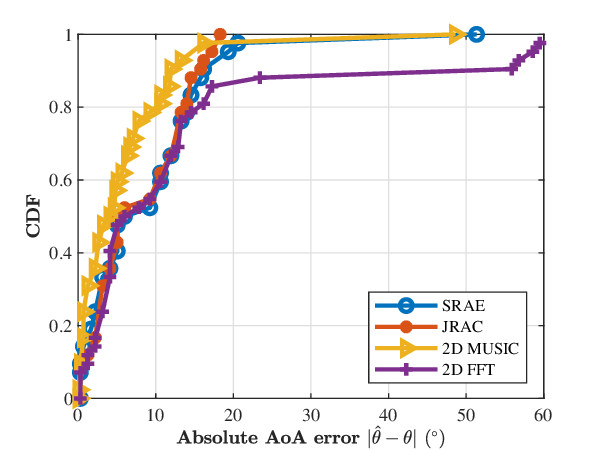}}\quad
\subfloat[Localization error, 4-tag setup\label{fig:usrp_loc_4_tags}]{\includegraphics[height=1.3in]{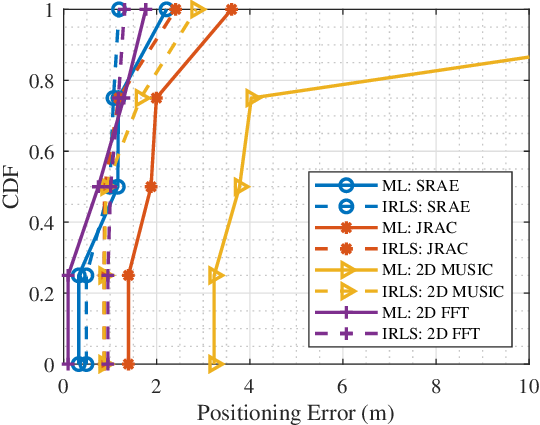}}\\
% \end{minipage}\quad
% \begin{minipage}{.3\linewidth}
  \subfloat[Range est. error, 100-tag setup\label{fig:usrp_range_est_100_tags}]{%
\includegraphics[height=1.3in]{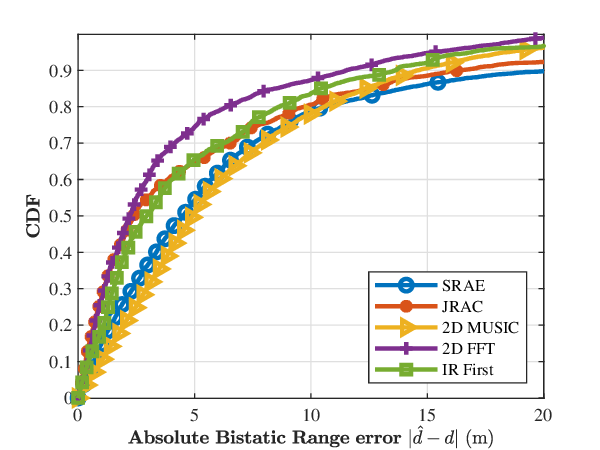}}\quad
\subfloat[Angle est. error, 100-tag setup\label{fig:usrp_ang_est_100_tags}]{%
\includegraphics[height=1.3in]{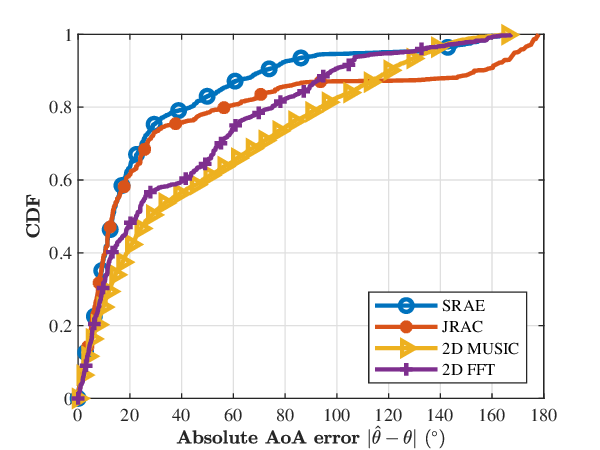}}\quad 
% \end{minipage}\quad
% \begin{minipage}{.3\linewidth}
\subfloat[Localization error, 100-tag setup\label{fig:usrp_loc_100_tags}]{\includegraphics[height=1.3in]{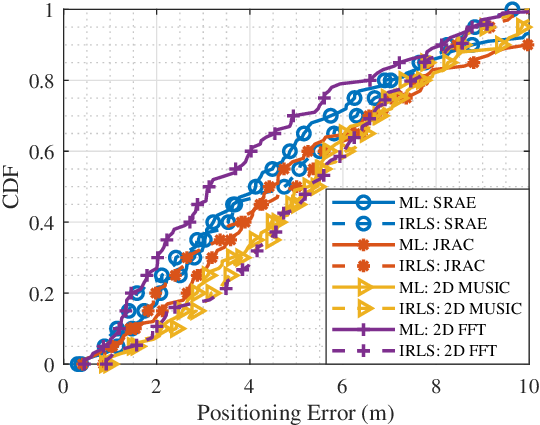}}
% \end{minipage}
  \caption{Joint range and angle estimation, and localization error on experimental setup with (a-c) 4 tags, and (d-f) 100 tags.
  (c,f) $A\!:\!B$ denotes the positioning method $A$ is applied to the ranges and AoAs which are  estimated by method $B$. ML method denotes ML-based Gradient Ascent with Line Search. 
}
% \label{fig:usrp_loc}
% \label{fig:usrp_est_4_tags}
% \label{fig:usrp_est_100_tags}    
\vspace{-.2in}
\end{figure*}

In \figref{fig:usrp_range_est_4_tags}, we show the CDF of absolute ranging error $|\widehat{d}-d|$ for the 4-tag scenario. The 2D FFT and IR First algorithms maintain performance similar to the simulated results in \figref{fig:sim-range-error}. In contrast, the 2D MUSIC, SRAE, and JRAC methods exhibit noticeable degradation in ranging accuracy. This degradation occurs because these algorithms rely on a MUSIC-based step for delay estimation.
MUSIC assumes a well-defined signal and noise subspace, which becomes unreliable in real-world multipath environments. Coherent multipath components and uncertainty in the number of subspaces significantly reduce the robustness of MUSIC for delay estimation, as noted in prior studies~\cite{shan1985spatialsmoothing,schmidt1986music}.

% In \figref{fig:usrp_range_est_4_tags}, we present the CDF of absolute ranging error $|\widehat{d}-d|$. While the 2D FFT and IR First maintains the performance similar to the simulated results in \figref{fig:sim-range-error}, 2D MUSIC, SRAE and JRAC methods observe degradation in the ranging performance. We believe this degradation is due to use of a MUSIC step in these algorithms. It is known fact that the use of MUSIC in delay estimation in real-world multipath environment has substantial limitations primarily due to known number of subspaces and lower performance in presence of coherent subspaces~\cite{}. 
In \figref{fig:usrp_ang_est_4_tags}, we present the CDF of absolute AoA estimation error. Unlike the range results, MUSIC-based algorithms outperform FFT-based methods for angle estimation. This improvement occurs because, at high SNR, spatial subspaces are more separable than delay subspaces, even in multipath environments. The ULA provides spatial diversity, which allows MUSIC to exploit subspace decomposition for high-resolution angle estimation. In contrast, FFT-based methods suffer from coarse grid resolution and spectral leakage, which limit angular accuracy. Therefore, while MUSIC struggles with delay estimation, it remains effective for AoA estimation under high-SNR conditions.

% \begin{figure}[tb]
%     \centering
%   \subfloat[Range estimation error.\label{fig:usrp_range_est_100_tags}]{%
% \includegraphics[width=0.7\columnwidth]{Results_USRP/range_CDF_plot_100tag_long.eps}}\\
% \subfloat[Angle estimation error\label{fig:usrp_ang_est_100_tags}]{%
% \includegraphics[width=0.7\columnwidth]{Results_USRP/AoA_CDF_plot_100tag_long.eps}}
%   \caption{Joint range and angle estimation errors on experimental setup with 100 tags.}
% \label{fig:usrp_est_100_tags} 
% \end{figure}

% \begin{figure}[tb]
%     \centering
%   \subfloat[CIRs from the 4-tag setup.\label{fig:usrp_CIR_example_4}]{%
% \includegraphics[width=0.8\columnwidth]{Results_USRP/CIR-4Tag.eps}}\\
% \subfloat[CIRs from the 100-tag setup.\label{fig:usrp_CIR_example_100}]{%
% \includegraphics[width=0.8\columnwidth]{Results_USRP/CIR-100Tag.eps}}
%   \caption{CIRs of carrier and backscatter channels from the 4-tag and 100-tag setups: The TX, RX and tags are located such that $d_0 = 8.6$m and $d_\back = 9.8$m. We manually add an offset of 10 m in the backscatter channel, thus, the ``distance" between LoS peaks in the channels is 11.2 m. Note that both channels are more cluttered in 100-tag setup compared to the 4-tag setup for the same SNR. The carrier channel with same SNR has wider peaks and bumps on the edge of those peaks indicating the impact of multipaths.}
% \label{fig:usrp_CIR_example} 
% \end{figure}

In \figref{fig:usrp_range_est_100_tags} and \ref{fig:usrp_ang_est_100_tags}, we present the CDF of absolute ranging and AoA errors in 100-tag setup. 
While the relative performance of various methods remain similar to that of the 4-tag setup, the overall performance of the estimation algorithms drops substantially. This is due to three possible reasons: (a) low SNR in 100-tag setup, (b) additional multipath introduced by the tags when not transmitting a packet. 
We observe that CIRs of the carrier and backscatter channels in the 100-tag setup are significantly more cluttered than in the 4-tag setup for the same TX, RX, and the tag. This shows that there are more multipaths in the environment with 100 tags, indicating that when the tags are on and not modulating a packet, they indeed act as (unmodulated) reflectors, due to the semi-passive architecture. Unlike passive tags where the tag antenna is connected to a matched load (and therefore absorbing incident RF signals) when not modulating, semi-passive tags have their antenna always connected to a mismatched load~\cite{ble2019tmtt}. Therefore, even when the tags are not modulating, the signals are reflected with a phase shift, resulting in additional multipaths. A potential resolution would be the use of multi-state RF front-ends to ``mute'' the tags that are not transmitting packets, by introducing a matched or low-reflection state (e.g. \cite{Kim14rftag}).
Furthermore, in \figref{fig:usrp_ang_est_100_tags}, we also observe the longer tail in the angle estimation error. The longer tails denote the fundamental limit of the angle estimation in the end-fire region of the ULA, in which many tags are located (refer to \figref{fig:usrp_map}). % which is highlighted in this graph.  from \figref{fig:usrp_CIR_example}

% In \figref{fig:usrp_range_est_100_tags}, we show the CDF of absolute ranging error for the dense 100-tag scenario. All algorithms experience a significant performance drop compared to the 4-tag case. This degradation results from increased packet collisions and lower SNR caused by simultaneous transmissions from many tags. Despite this, the relative ranking of algorithms remains consistent: 2D FFT and IR First outperform MUSIC-based methods in range estimation.

% In \figref{fig:usrp_ang_est_100_tags}, we show the CDF of absolute AoA estimation error for the 100-tag scenario. The overall performance deteriorates due to the same factors-—collisions and reduced SNR. Additionally, we observe longer tails in the error distribution. These tails occur because many tags lie in the end-fire region of the RX array, where angular resolution is inherently poor. This limitation is a fundamental property of ULAs and highlights the geometric constraints of array-based AoA estimation. 
\begin{table}[tb]
\centering
\scalebox{0.9}{%
\begin{minipage}{1\linewidth}
\caption{Computational complexity and Runtime of Joint Range and Angle Estimation Methods for $N_\sym=98$ OFDM symbols.} %captured from the experimental setup.}
\label{tab:realtime_complexity_joint_range_aoa}
\centering
\begin{tabular}{c|c|c} 
\hline
\begin{tabular}[c]{@{}c@{}}\textbf{Estimation}\\\textbf{Method}\end{tabular} & \begin{tabular}[c]{@{}c@{}}\textbf{Computational}\\\textbf{Complexity}\end{tabular} & 
\begin{tabular}[c]{@{}c@{}}\textbf{Average}\\\textbf{Runtime (ms)}\end{tabular} \\\hline
SRAE & $\mathcal{O}(G_{\tau}N_\subc^3+G_{\theta}N_\ant^3)$ & 55  \\\hline
JRAC & $\mathcal{O}(G_{\tau}N_\subc^3+G_{\tau}G_{\theta}N_\subc N_\ant)$ & 475  \\\hline
2D MUSIC & $\mathcal{O}(G_\tau G_\theta N_\ant^2 N_\subc^2+N_\ant^3 N_\subc^3)$ & 10380 \\\hline
2D FFT & $\mathcal{O}(G_\tau G_\theta \log (G_\tau G_\theta))$ & 2038 \\\hline
\end{tabular}
\end{minipage}}
\end{table}
\tabref{tab:realtime_complexity_joint_range_aoa} presents the average runtime required to compute a range and angle estimate from each tag packet using the four algorithms. The SRAE and JRAC methods outperform the others by orders of magnitude, achieving near real-time processing capability. In contrast, 2D FFT and 2D MUSIC incur significantly higher computational costs due to their reliance on complex FFT and expensive grid searches, respectively. This difference in runtime becomes critical for scalability in dense deployments. 
While the relative accuracy and the runtime of different range-angle estimation algorithms varies, ultimately the metric of interest is the overall positioning quality using the respective range and angle estimates.%, as discussed in the next section.
%\vspace{-.15in}
% \begin{figure}
%     \centering
%     \subfloat[4 Tags\label{fig:usrp_loc_4_tags}]{\includegraphics[width=.7\columnwidth]{Results_USRP/CDF-Localization-4-tags-wGradient.eps}}\\
%     \subfloat[100 Tags\label{fig:usrp_loc_100_tags}]{\includegraphics[width=.7\columnwidth]{Results_USRP/CDF-Localization-100-tags-long-wGradient.eps}}
%     \caption{CDF of localization error on experimental setup: (a) With 4 tags, (b) With 100 tags. The ML method indicates ML-based Gradient Search with Line Search.$A\!:\!B$ denotes that the positioning method $A$ is applied to the ranges and  AoAs estimated by method $B$.
% }
%     \label{fig:usrp_loc}
% \end{figure}

\begin{table}[tb]
\centering
\scalebox{0.9}{%
\begin{minipage}{1\linewidth}
\caption{Runtime of Positioning Methods}
\label{tab:realtime_complexity_positioning}
\resizebox{\columnwidth}{!}{
\begin{tabular}{c|c|c|c} 
\hline
\multirow{2}{*}{\begin{tabular}[c]{@{}c@{}}\textbf{Positioning}\\\textbf{Method}\end{tabular}} & \multirow{2}{*}{\begin{tabular}[c]{@{}c@{}}\textbf{Computational }\\\textbf{Complexity}\end{tabular}} & \multicolumn{2}{c}{\textbf{Runtime (ms)}} \\ \cline{3-4}
 &  & \textbf{4-tag} & \textbf{100-tag} \\\hline
ML Grid Search & $\mathcal{O}(D(M_\text{r} + M_\angle) \varepsilon_\text{ML}^{-D})$ & 4670 & 22802 \\\hline
\begin{tabular}[c]{@{}c@{}}ML Gradient Ascent~\\with Line Search\end{tabular} & $\mathcal{O}(D(M_\text{r} + M_\angle)M_\text{r} |\calS| K_\text{ML})$ & 500 & 15818 \\\hline
IRLS & $\calO(DK(M_\text{r} + DM_\angle)^2)$ & \textbf{38} & \textbf{51} \\\hline
\end{tabular}}
\end{minipage}
}
\end{table}
%\subsection{Positioning Performance}
%--------------------------
We evaluate the positioning performance of two optimization methods described in \secref{sec:positioning}:
(a) ML-based Gradient Ascent with Line Search proposed in Algorithm~\ref{alg::alg-gd} with $\sigma_n,\kappa_n = 1$ and the line search step sizes $\calS=\{0.001, 0.01, 0.1, 1\}$. (b) IRLS proposed in \secref{sec::irls} with $\varepsilon_{\text{IRLS}} = 10^{-6}$. 
The analysis uses range-angle estimates obtained from four algorithms: JRAC, SRAE, 2D MUSIC, and 2D FFT.%, with parameters described in \secref{sec:usrp_est}.
\figref{fig:usrp_loc_4_tags} and \ref{fig:usrp_loc_100_tags} shows the CDF of positioning error for the 4-tag and 100-tag datasets. The 4-tag dataset achieves substantially better positioning accuracy than the 100-tag dataset. This improvement is expected because the 4-tag scenario represents a simplified environment with high SNR, whereas the 100-tag scenario introduces lower SNR and heavy multipath, which degrade localization accuracy. 
Among the range-angle estimation algorithms, 2D FFT combined with ML-based positioning achieves the lowest positioning error. This result aligns with the robustness of 2D FFT on real-world data and the optimality of ML-based estimation. Interestingly, the two low-complexity algorithms proposed in this paper--JRAC and SRAE--deliver positioning accuracy comparable to 2D FFT when paired with IRLS. This finding demonstrates that reducing computational complexity in range-angle estimation does not necessarily compromise positioning quality, making JRAC and SRAE practical for large-scale deployments.

\tabref{tab:realtime_complexity_positioning} reports the average runtime for computing position estimates using ML Gradient Ascent with Line Search and IRLS for both datasets. It also includes a brute-force baseline that searches the ML objective over a grid with 5 cm granularity. The results show that ML-based methods scale with the number of tags and the number of measurements per tag, both of which are higher in the 100-tag dataset. In contrast, IRLS requires only the inversion of a tall matrix with fixed rank, so its runtime grows much more slowly. Overall, ML with gradient ascent demands significantly more computational time: at least 10$\times$ higher than IRLS in the 4-tag setup and 300$\times$ higher in the 100-tag setup. These results highlight that while ML and IRLS achieve comparable positioning accuracy, IRLS offers a clear advantage in scalability.

Overall, even though the ranking of range-angle estimation algorithms differs between range and angle accuracy, the ultimate metric of interest is positioning quality. Our results show that JRAC and SRAE, combined with IRLS, deliver the best trade-off between accuracy and computational efficiency. This makes them strong candidates for practical multi-static localization systems.
\section{Conclusion}
In this work, we introduced the JRAC and SRAE methods for real-time joint range–AoA estimation in multistatic BNs. These two methods achieve range and angle measurement accuracy comparable to FFT- and subspace-based baselines, while significantly reducing computational complexity.   Building on these measurements, we investigated tag localization using both ML and IRLS algorithms. Our experimental results revealed that an unmodified IRLS approach, as commonly adopted in prior literature, can assign artificially high weights to outlier measurements, leading to degraded localization performance. To address this limitation, we proposed a scaling scheme that effectively mitigates the impact of outliers. We further demonstrated that the scaled IRLS algorithm achieves localization accuracy comparable to brute-force search  on the ML criterion, while significantly reducing computational complexity. This reduction enables practical real-time tag localization, making the proposed approach well suited for use in large-scale, time-critical BN deployments.
\vspace{-0.1in}
\bibliographystyle{IEEEtran}
\bibliography{IEEEabrv,refs}%,scalability_references}
%\onecolumn
%\include{appendix}
\end{document}